\documentclass[prb,aps,twocolumn,showpacs]{revtex4-1}
\usepackage{graphicx}
\usepackage{epstopdf}
\usepackage{amsmath,amssymb,dsfont,upgreek}
\usepackage{natbib}
\usepackage[colorlinks=true,  linkcolor=blue, citecolor=blue, urlcolor=blue,hyperfootnotes=false]{hyperref}
\usepackage{soul}
\begin{document}
\date{\today}

\title{Excitonic instabilities and spontaneous time-reversal symmetry breaking on the Honeycomb lattice}

\author{Wei Liu}
\affiliation{Physics Department, City College of the City University of New York, New York, NY 10031, USA}
\author{Alexander Punnoose}%
\email{Corresponding Author: punnoose@ift.unesp.br} \affiliation{Instituto  de F\'\i sica Te\'orica - Universidade Estadual Paulista, R.\ Dr.\ Bento Teobaldo Ferraz 271, Barra Funda, S\~ao Paulo - SP, 01140-070, Brazil}

\begin{abstract}
We elucidate the close relationship between spontaneous  time-reversal symmetry breaking and the physics of excitonic instabilities in strongly correlated multiband systems. The underlying mechanism responsible for the spontaneous breaking of time-reversal symmetry in a many-body system is closely related to the  Cooper-like pairing instability of \textit{interband} particle-hole pairs involving higher order symmetries. Studies of such pairing instabilites have, however, mainly focused on the mean-field aspects of the virtual exciton condensate, which ignores the presence of the underlying collective Fermi liquid excitations.  We show that  this relationship can be exploited to systematically derive the coupling of the condensate order-parameter  to the \textit{intraband} Fermi liquid particle-hole excitations. Surprisingly, we find that the \textit{static} susceptibility  is negative in the ordered phase when  the coupling to the Fermi liquid collective excitations are included, suggesting that  a uniform condensate of virtual excitons, with or without time reversal breaking, is an unstable phase at $T=0$.
\end{abstract}

%There is considerable interest in exploring time-reversal symmetry broken phases driven by their proposed relevance to high temperature superconductivity.


\maketitle
\section{Introduction}
\begin{figure}[t]
\includegraphics[width=0.6\linewidth]{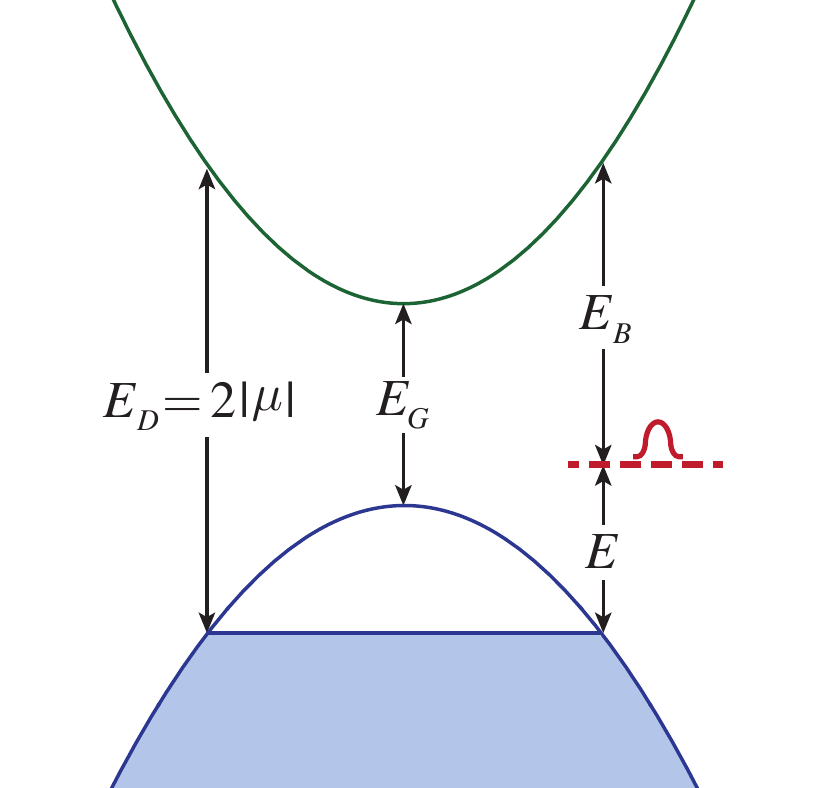}
\caption{%\textcolor{red}{[The arrows are not the same size. Lower $\mu$ and bring it closer to the Fermi energy. Draw a gaussian  at $E$.]}
A simple direct bandgap ($E_G$)  model with quadratic bands of equal masses  is shown. The location of the exciton level $E=E_D-E_B$ is marked by the dashed line.  $E_B$ is the exciton binding energy  and $E_D$ is the direct interband excitation  threshold from the chemical potential $\mu$ (measured from the center of the gap).   When $E_B=E_D$, the cost of creating virtual exciton pairs become negligible. }  \label{fig:boundstate}
\end{figure}

In multiband systems, the normal Fermi liquid (FL) state of a partially filled band becomes unstable when the exciton~\cite{Mott_Transition} (electron-hole pair) binding energy, $E_B$,  is tuned to become larger than the direct excitation threshold energy $E_D<|E_B|$ (see Fig.\,\ref{fig:boundstate}).  When this happens  \textit{virtual} excitons form spontaneously in the groundstate and destabilizes the FL state. As a cure for this instability, it was proposed many decades ago~\cite{Knox, Cloizeaux} that a new Hartree-Fock groundstate may be constructed by hybridizing the bands.  Such a state is characterized by a non-zero expectation value $\langle a_\mathbf{k}^\dagger b_\mathbf{k}\rangle\neq 0$, where $a^\dagger_\mathbf{k}/a_\mathbf{k}$ and $b_\mathbf{k}^\dagger/b_\mathbf{k}$ are the electron creation/annihilation operators in the relevant bands. As a result, the relative phase between the bands is locked to form a uniform ($q=0$) condensate of interband particle-hole pairs. %We examine this   proposal  in detail in this paper.

To distinguish condensates of \textit{interband} particle-hole pairs from quantum liquids  that arise from instabilities in the \textit{intraband} particle-hole channel found in single band systems, like the Pomeranchuk instability~\cite{pomeranchuk},  we refer to the former here as an ``excitonic-liquid" (XL).

The XL is to be contrasted with the much studied \textit{excitonic-insulator}  state  formed when the number of free conduction electrons and valence holes are finite. The latter can occur either naturally  when the bands  cross at the Fermi level (semimetal  with bandgap $E_G<0$) or  can be created artificially by  optically pumping a semiconductor ($E_G>0$).  The two carrier types have their own chemical potentials which are tied together by thermodynamic considerations. The excitonic-insulator problem has been shown to be mathematically equivalent to the BCS (Bardeen-Cooper-Schrieffer) model with  electron-hole pairs instead of electron-electron pairs~\cite{Keldysh_Kopaev, Kozlov_Maksimov, jerome_rice_kohn, RMP_Halperin_rice, Nozieres_Comte}.
Unlike the gapped  excitonic-insulator state, the XL state has only one species of free carriers, hence away from half-filling the chemical potential lies in the band leading to a  metallic state with a sharp Fermi surface (at least in the Hartree-Fock approximation). This raises the  interesting question of the stability of the XL in the presence of  gapless fermions at the Fermi surface. %We address this issue in detail  in this paper.

The issue of the coupling of  bosonic modes in the ordered phase to gapless fermions has received considerable attention recently, particularly concerning the applicability of the Hertz theory~\cite{hertz}  to continuous quantum phase transitions at $T=0$. The issue is that integrating the fermions out completely has been shown to give rise to a non-local (in time or frequency) bosonic theory that is dominated by Landau damping, which in most cases lead to an infinite number of marginal terms in the effective action~\cite{abanov_chubukov}.  (For a review of this and other related issues, see Ref.~\onlinecite{RMP_hertz_breakdown}.)  In all these studies, involving a single band, it is usually implicitly assumed, to ensure  the stability of the condensate, that the opposite limit, i.e., the static limit,  of the fluctuation of the order-parameter  $\lim_{q\rightarrow 0}\langle |\delta \phi(q,\omega=0)|^2\rangle >0$ is positive.
We find that this assumption  is generally violated  in a multi-band system when the coupling of the order-parameter of the XL state to the collective intraband particle-hole excitations is considered. We note that the issues originating from the dynamics of the gapless fermions leading to the Landau damping of the bosonic mode is irrelevant for us in establishing the stability of the condensed phase.

There is considerable interest to understand if a \textit{uniform} XL can in principle exist. In fact,  excitonic singularities in multiband systems is recognized to be one of the very few known mechanisms capable of introducing singularities in the irreducible interactions leading to the breakdown of  FL theory~\cite{varma_SFL_review}.
At the mean-field level, the Hamiltonian couples linearly to the bilinear operators $a^\dagger_\mathbf{k}b_\mathbf{k}$ via a vertex function $\mathcal{A}(\mathbf{k})$. The symmetries of $\mathcal{A}(\mathbf{k})$ are dictated by the lattice symmetries. As a result, a non-zero expectation value
$\mathcal{A}(\mathbf{k})\langle a^\dagger_\mathbf{k}b_\mathbf{k}\rangle\neq 0$  partially breaks  the point group symmetry. Furthermore, depending on the transformation properties of $\mathcal{A}(\mathbf{k})$  under the corresponding magnetic group, time-reversal ($\mathcal{T}$) symmetry  may or may not be spontaneously broken in the groundstate. %~\cite{Varma_1997}.
%Hence, the XL states with different symmetries characterized by $\mathcal{A}(\mathbf{k})$ provide a natural basis to classify  strongly correlated many-body groundstates in multiband systems with excitonic instabilities~\cite{}.
%
Special interest in understanding the properties of electronic systems with broken $\mathcal{T}$ symmetry %, but translationally invariant, groundstates
stem from its proposed relevance to high temperature superconductivity. An intriguing proposal, put forward originally by Varma as a candidate for a Marginal Fermi liquid, posited a non-magentic, translationally invariant groundstate with spontaneously broken  $\mathcal{T}$ invariance\,\cite{Varma_1997}.

More generally, such groundstates are novel in that they can exhibit anomalous Hall and  Kerr effects without  magnetic fields.  Various non-magnetic, translationally invariant states with broken $\mathcal{T}$  symmetry that are ubiquitous in Hartree-Fock approximation schemes have been classified and analyzed in Ref.\,\onlinecite{Fradkin_sun_TRV}.  They were first demonstrated by Haldane in a simple two-dimensional model  of non-interacting electrons on the honeycomb lattice with a periodic magnetic field arranged in such a way that the field averages to zero in each unit cell thus preserving the translational symmetry of the lattice\,\cite{haldane_graphene_TRV}.  It was later shown  that interacting electrons on the honeycomb lattice with next-nearest neighbor interactions  can form exactly the same  state as proposed by Haldane that  spontaneously breaks $\mathcal{T}$ invariance  in the groundstate\,~\cite{Zhang_TMI}. %The stability of such a state to doping is studied in this paper. % in this paper. %It is important to note that the breaking of $\mathcal{T}$ invariance  in the above works are not related to the spin of the electrons.

%Since $\mathcal{A}(\mathbf{k})$ can be thought of as the wave function of a single exciton~\cite{jerome_rice_kohn},  the $\mathcal{T}$ symmetry broken states may be classified by the symmetry of the uniform XL state. %Depending on the nature of the $\mathcal{T}$ symmetry breaking,

Our main purpose in this paper is to analyze the stability of the mean-field solutions of particle-hole condensates both with and without $\mathcal{T}$ symmetry breaking.  We first show that the XL's are metallic \textit{away} from half-filling. Indeed, since the single-particle states in any mean-field  approximation of a particle number conserving condensate  remains in one-to-one correspondence with the FL states with well-defined quasiparticle excitations, at the Hartree-Fock level  the condensates away from half-filling are always metallic with a continuum of gapless fermionic excitations at the Fermi surface.

We show that the hybridization of the bands necessarily induces a coupling of the XL order-parameter  to the gapless fermions at the Fermi surface. In the following, we analyze the static susceptibility originating from this coupling in two different two-band models, both of which stabilizes an XL phase at the mean-field level. % into an XL phase. % beyond some critical strength of the particle-hole interaction.
The first model is a simplified continuum degenerate semiconductor model  in which the $\mathcal{T}$ symmetry is left intact in the condensed phase,  and the second model is a lattice model  that supports a $\mathcal{T}$ symmetry broken  phase.

\section{Theory of the XL state.}
Our objective in this section is to analyze a simple model that undergoes a quantum phase transition to  a  uniform XL state. We consider the well studied, continuum,  two-band, spinless fermion model with a direct bandgap~$E_G>0$ shown in Fig.\,\ref{fig:boundstate}. Keeping only  the direct intraband scattering of the electrons and holes, the model is described by the Hamiltonian
\begin{equation}
K=\sum_{\mathbf{k}}\psi^\dagger_\mathbf{k}\left(\epsilon_\mathbf{k}
{\uptau}_3-\mu\right)\psi_\mathbf{k} -\frac{1}{L^{d}}\!\!\sum_\mathbf{k',k;q}V_{\mathbf{kk'}}
a^\dagger_{\mathbf{k+q}}b_\mathbf{k}b^\dagger_{\mathbf{k'}}a_\mathbf{k'+q}
\label{eqn:H}
\end{equation}
The $2\times 2$ Pauli matrices $\uptau_i$'s act on the internal band space;  ${\uptau}_3$  is the $z$-component. The spinor $\psi^\dagger_\mathbf{k}=(b^\dagger_\mathbf{k},a^\dagger_\mathbf{k})$ creates \textit{electrons}  in the conduction ($b$) and the valence ($a$) bands. The bands are assumed to be parabolic  with equal masses   $\epsilon_\mathbf{k}={E_G}/{2}+k^2/2m$. The chemical potential $\mu$ constraints the  electron number ${N}=\sum_\mathbf{k}\psi^\dagger_\mathbf{k}\psi_\mathbf{k}$. Hole doping is assumed throughout, i.e., partially filled valence band and empty conduction band, hence $\mu<0$. All energies are measured from the center of the gap.

For the interaction, only the dominant scattering channel, the direct intraband scattering of the electrons and holes, are retained. The Hamiltonian $K$  has been used extensively  as a minimal model to analyze the excitonic properties  of optically pumped degenerate semiconductors~\cite{Nozieres_Gavoret, Nozieres_Comte,Nozieres_Comte_2}. We note that because optical pumping  fixes the number of electrons and holes ${N}_{ph}=\sum_\mathbf{k}b^\dagger_\mathbf{k}b_\mathbf{k}+a_\mathbf{k}a^\dagger_\mathbf{k}$, the chemical potential, $\mu_{ph}$, in these systems  appear as~\cite{Nozieres_Comte,cote_griffin}  $K_{ph}=H-\mu_{ph}N_{ph}= \sum_\mathbf{k}\psi_\mathbf{k}^\dagger
[(\epsilon_\mathbf{k}-\mu_{ph})\uptau_3]\psi_\mathbf{k}+H_\textrm{int}$.
A  particle-hole transformation, $a_\mathbf{k}\rightarrow c^\dagger_{-\mathbf{k}\downarrow}$ and $b_\mathbf{k}\rightarrow c_{\mathbf{k}\uparrow}$,  maps  $K_{ph}$ to the continuum fermion model with attractive interactions~\cite{randeria_BCS_BEC, BCS_phase_Loktev_review} whose  mean-field solution   corresponds exactly to the  BCS Hamiltonian. In contrast, the same transformation maps $K$ in Eq.\,(\ref{eqn:H}) to the BCS Hamiltonian in a Zeeman field with field $h_Z=\mu$  and fixed  BCS chemical potential  $\mu_{BCS}=-E_G/2$. When $E_G<0$ (semi-metal)  the mean-field solution of $K$  maps exactly to the BCS Hamiltonian in a Zeeman field~\cite{varma_balents_X_PRL, X_Gorkov}.  Since $E_G>0$ when the XL state is formed, the particle-hole transformation does not provide much further insights. However, we show that the instability is closely related to the Cooper-like pairing  of interband particle-hole pairs at the Fermi surface.

%\footnote{We assume a Jellium model for the positive background, hence reordering the electron operators, which is responsible for the negative sign in $H$ in Eq.~(\ref{eqn:H}),  generates terms proportional to  $V(0)$ which are exactly canceled by the background energy. Hence we set $V(0)=0$.}.
%

Further analytical progress is possible by restricting our analysis to the $s$-wave pairing state. We make the standard approximation of  a  separable  screened potential\,\cite{Nozieres_Gavoret},
%
%\begin{eqnarray}
$V_\mathbf{kk'}=V\mathcal{A}(\mathbf{k}) \mathcal{A}(\mathbf{k'})$,  where $\mathcal{A}(\mathbf{k}) =1$ for $|-\epsilon_\mathbf{k}-\mu| < \Lambda$ and zero otherwise (around the hole Fermi surface). The cutoff is  of the order $\Lambda\sim \mathcal{O}(E_G)$.
%
%u_k&=&\left\{\begin{array}{ccc}
%1 & \textrm{if}&|\epsilon_\mathbf{k}-\mu| <\Lambda\\
%0 & \textrm{if}&|\epsilon_\mathbf{k}-\mu| >\Lambda\label{eqn:cutoff}
%\end{array}\right.
%\end{eqnarray}
%
%Also, after an appropriate particle-hole transformation, the model transforms into the BCS Hamiltonian in a parallel magnetic field.
%
%Furthermore, consistent with our ignoring the spin degrees of freedom, the  interband electron-hole exchange is not considered in this model%; hence, $\mathcal{T}$ breaking effects due to the weak ferromagnetism proposed in Ref.\,\textcolor{red}{[Volkov]} are not observed in this model.
%
This simplification allows the interaction  to be written as a product of bilinear operators
\begin{equation}
H_\textrm{int}=-\frac{V}{L^d}\sum_\mathbf{q}\hat{\Phi}^\dagger(\mathbf{q})\hat{\Phi}(\mathbf{q})
\label{eqn:Hint}
\end{equation}
where,
%\begin{equation}
$\hat{\Phi}(\mathbf{q})=\sum_\mathbf{k}\nolimits'\psi^\dagger_\mathbf{k}
\uptau^+\psi_{\mathbf{k+q}}
=\sum_\mathbf{k}\nolimits' b^\dagger_{\mathbf{k}}a_{\mathbf{k+q}}$,
%\end{equation}
%
and $\uptau^+$ is the band index raising operator. The prime on the $k$-sum explicitly enforces the restriction on $\mathcal{A}(\mathbf{k})$. Thus we arrive at our simplified model Hamiltonian
\begin{equation}
K=\sum_{\mathbf{k}}\psi^\dagger_\mathbf{k}\left[\epsilon_\mathbf{k}
{\uptau}_3-\mu\right]\psi_\mathbf{k} -\frac{V}{L^d}\sum_\mathbf{q}\hat{\Phi}^\dagger(\mathbf{q})\hat{\Phi}(\mathbf{q})
\label{eqn:K}
\end{equation}

\subsection{Fermi surface instability: Mean-field theory}
We explore the possibility of a uniform  XL groundstate   with a finite  expectation value $V\langle\hat{\Phi}(\mathbf{q})\rangle/L^d=\phi_0\delta_{q,0}$. Such a state is anticipated from the behavior of the effective interaction, $\mathcal{U}(\mathbf{q})=V\Gamma(\mathbf{q})$, which in the ladder diagram approximation  takes the general form
\begin{equation}
\Gamma(\mathbf{q})=\frac{1}{1-V\chi_0(\mathbf{q})}
\label{eqn:Uq}
\end{equation}
It follows from the form of  $K$ in Eq.\,(\ref{eqn:K}) that $\chi_0$ is the non-interacting interband susceptibility
\begin{equation}
\chi_0(\mathbf{q})=-\frac{1}{L^d}\sum_\mathbf{k}\nolimits'
\frac{f(\epsilon_\mathbf{k+q}^+)-f(\epsilon_\mathbf{k}^-)}
{\epsilon^+_\mathbf{k+q}-\epsilon^-_\mathbf{k}}
\label{eqn:chi0}
\end{equation}
 The energies $\epsilon^\pm_\mathbf{k}=\pm \epsilon_\mathbf{k}-\mu$ are the upper and lower band energies measured from $\mu$.  Since we assume hole doping,  at $T=0$  the Fermi function  $f(\epsilon_\mathbf{k}^+)=0$ for all $\mathbf{k}$, and $f(\epsilon_\mathbf{k}^-)=1$  for $|\mathbf{k}|>k_F$ and  zero otherwise. $k_F$ is the non-interacting Fermi wavevector that fixes the filling. Substituting for the Fermi functions in (\ref{eqn:chi0}), we get  $\chi_0(0)>0$ in the limit $q=0$. Hence the uniform $\Gamma(0)$ diverges  at a critical $V_c$  satisfying $1-V_c\chi_0(0) = 0$, indicating a FL instability for the states at the chemical potential. The critical $V_c$ that determines the QCP equals
\begin{equation}
\frac{1}{V_c}=\sum_\mathbf{k}\nolimits'' \frac{1}{2\epsilon_\mathbf{k}}
\label{eqn:Vc}
\end{equation}
[The sum $\sum_\mathbf{k}\nolimits''={L^{-d}}\sum_{|\mathbf{k}|>k_F}\nolimits'$. (Remember that the single prime denotes the upper cutoff imposed by $\mathcal{A}(\mathbf{k})$.)]

%\subsection{\textbf{Binding energy of a single exciton}}
%\label{sec:cooper-pair}

We show below, using a more elementary quantum mechanical calculation, that the above instability is related to the formation of Cooper-like pairs of interband particle-hole pairs with zero energy. This is done by writing the equation of motion for a single particle-hole pair in the background of a `rigid' Fermi sea of particles~\cite{jerome_rice_kohn}.

We assume a linear combination of pair wavefunctions, $\varphi_\mathbf{k}$, of single particle-hole pairs created by annihilating  an electron from inside of the Fermi surface, $|FS\rangle$, of the partially filled valence band  and  recreating it at the same $\mathbf{k}$ value in the empty conduction band
\begin{equation}
|\phi_0\rangle=\sum_\mathbf{k}\nolimits''\varphi_\mathbf{k} b^\dagger_\mathbf{k}a_\mathbf{k}|FS \rangle
\end{equation}
The sum extends from $k_F$  to the upper cutoff.

The pair energy is obtained by solving the stationary  Schr\"{o}dinger equation
\begin{eqnarray}
&&E\sum_\mathbf{k}\nolimits''\varphi_\mathbf{k}\, b^\dagger_\mathbf{k}a_\mathbf{k} = \sum_\mathbf{k}\nolimits''\varphi_\mathbf{k}
\left[K,b^\dagger_\mathbf{k}a_\mathbf{k}\right]\\
&=&\sum_\mathbf{k}\nolimits'' \left[ 2\epsilon_\mathbf{k}\varphi_\mathbf{k}-\frac{V}{L^d}\sum_\mathbf{k'}\nolimits'\varphi_\mathbf{k'}(a^\dagger_\mathbf{k'}a_\mathbf{k'}-b^\dagger_\mathbf{k'}b_\mathbf{k'})\right]b^\dagger_\mathbf{k}a_\mathbf{k}\hspace{0.5cm}
\end{eqnarray}
The Cooper model assumes a rigid Fermi sea, which is equivalent to decoupling the interaction terms by averaging over the Fermi sea $\langle a^\dagger_\mathbf{k'}a_\mathbf{k'}-b^\dagger_\mathbf{k'}b_\mathbf{k'}\rangle=f(\epsilon^-_\mathbf{k'})-f(\epsilon^+_\mathbf{k'})$.   One then obtains the Bethe-Goldstone equation:
\begin{equation}
E\varphi_\mathbf{k}=2\epsilon_\mathbf{k}\varphi_\mathbf{k}-V\sum_\mathbf{k'}\nolimits''\varphi_\mathbf{k'}
\label{eqn:pairwavefunction}
\end{equation}
Substituting $\phi_0=V\sum_\mathbf{k}\nolimits''\varphi_\mathbf{k}$ and inverting the above equation gives the self-consistent equation
\begin{equation}
1=-V\sum_\mathbf{k}\nolimits''\frac{1}{E-2\epsilon_\mathbf{k}}
\end{equation}
Setting $E=0$ gives Eq.\,(\ref{eqn:Vc}). Since $E$ is measured from the chemical potential, when $V>V_c$, it becomes favorable to create particle-hole pairs out of the Fermi sea thus destabilizing the normal groundstate.

A possible cure for this instability is to assume a new Hartree-Fock solution built of virtual excitons\,\cite{Knox,Cloizeaux}. This is equivalent to assigning a non-zero expectation value to $\langle\hat{\Phi}(\mathbf{q})\rangle\sim\phi_0\delta_{q,0}$.
We note that the order-parameter is not invariant under a global $U(1)$ rotation $e^{i\frac{\theta}{2} \uptau_3}\psi_\mathbf{k}$ of the  relative phase between the two bands. However,  $H_\textrm{int}$ in Eq.\,(\ref{eqn:Hint}) conserves the particle number in each band separately and is therefore invariant. Hence the condensation spontaneously breaks the $U(1)$ symmetry. The associated Goldstone mode is easily identified with the phase of the XL order-parameter.  We note, however, that the enhanced $U(1)$ symmetry of $H_\textrm{int}$  exists  because terms such as $a^\dagger a^\dagger b b$ and $a^\dagger a^\dagger a b$ that do not conserve the particle number in each band are omitted in our model. It is  therefore only  an approximate symmetry in general and when spontaneously broken will give rise to a \textit{massive} pseudo-Goldstone mode~\cite{NJL_1,NJL_2}. For the sake of generality,  we suppress the gapless phase fluctuations in our simplified model by assuming $\phi_0$ to be real.

We proceed to look for a mean-field solution by substituting $\phi_0$ in Eq.\,(\ref{eqn:K}). The mean-field Hamiltonian
\begin{equation}
K_\textrm{MF}=\sum_\mathbf{k}\nolimits' \psi^\dagger_\mathbf{k}\left[\epsilon_\mathbf{k}\uptau_3-\mu-\phi_0\uptau_1\right]\psi_\mathbf{k}
+ \frac{L^d}{V}\phi_0^2
\label{eqn:K_MF}
\end{equation}
Since $\phi_0$ is assumed real, the linear terms in  $K_\textrm{MF}$  that originate on decoupling the interaction terms are combined as $\phi_0 (\hat{\Phi}^\dagger(0)+\hat{\Phi}(0))=\phi_0 {\psi}_\mathbf{k}^\dagger\uptau_1\psi_\mathbf{k}$. [From Eq.\,(\ref{eqn:K}),  the $q=0$ component $\hat{\Phi}(0)=\sum_\mathbf{k}\nolimits'\psi^\dagger_\mathbf{k}
\uptau^+\psi_{\mathbf{k}}
=\sum_\mathbf{k}\nolimits' b^\dagger_{\mathbf{k}}a_{\mathbf{k}}$.]
Diagonalizing $K_\textrm{MF}$, we obtain
%\label{eqn:Kmf}
%\end{equation}
%
%
%
$ \xi^\pm_\mathbf{k}=\pm E_\mathbf{k}-\mu$, where the energy $E_\mathbf{k}= \sqrt{\epsilon_\mathbf{k}^2+\phi_0^2}$. When $E_G>0$,  the  levels never cross~%
and all states up to the non-interacting  Fermi wavevector $k_F$ are filled. This point is worth emphasizing again, which is that because $K_\textrm{MF}$ commutes with the individual number operator $\hat{n}_\mathbf{k}=a^\dagger_\mathbf{k}a_\mathbf{k}
+b^\dagger_\mathbf{k}b_\mathbf{k}$, every $\mathbf{k}$-state up to the non-interacting $k_F$ remains occupied as $V$ is tuned through the QCP. It implies the  Fermi wavevector $k_F$ is not renormalized and a sharp Fermi surface exists in the condensed phase.

%
%\textcolor{red}{(**** Delete this section ***)}
%
 The most suggestive way to write the XL wavefunction, $|XL \rangle$, so that $k_F$ and the exciton nature are both apparent is  $|XL\rangle = \prod_\mathbf{k} [u_\mathbf{k}+ v_\mathbf{k} b^\dagger_\mathbf{k}a_\mathbf{k}] |FS \rangle$. The `vacuum'  is the non-interacting  Fermi surface $|FS\rangle = \prod_{\mathbf{|k|}>k_F}\nolimits'a^\dagger_\mathbf{k}|0\rangle$ corresponding to the partially filled valence band. %(Remember that the valence band is inverted and hence the  occupied states correspond to  $\mathbf{|k|}>k_F$.)
 Despite the formal similarity with the BCS wavefunction, it can be shown explicitly that $|XL\rangle $ does not possess off-diagonal long-ranged order (ODLRO) \cite{jerome_rice_kohn} due to  the sharp cutoff at $k_F$. In standard notation, the hybridized states are written as $c_{\mathbf{k}+}= u_\mathbf{k} b_\mathbf{k}-v_\mathbf{k}a_\mathbf{k} $ and $c_{\mathbf{k}-}=u_\mathbf{k}a_\mathbf{k} +  v_\mathbf{k} b_\mathbf{k}$, corresponding to $\pm E_\mathbf{k}$, respectively. The coefficients are found by minimizing the free energy. Setting all the phases to zero, which is justified when $\phi_0$ is real, we get
\begin{equation}
u_\mathbf{k}^2=\frac{1}{2}\left(1+\frac{\epsilon_\mathbf{k}}{E_\mathbf{k}}\right), \hspace{0.05\linewidth}
v_\mathbf{k}^2=\frac{1}{2}\left(1-\frac{\epsilon_\mathbf{k}}{E_\mathbf{k}}\right)
\label{eqn:uv}
\end{equation}

\subsection{Order-parameter fluctuations}

The existence of a sharp Fermi surface in the XL phase implies that gapless particle-hole excitations exists at the Fermi surface, which can couple to the fluctuations of the order-parameter of the XL.
The fluctuations about the mean-field solution are most easily calculated  using the standard functional integral method~\cite{book_popov,book_kleinart}. Briefly, the partition function %$Z=\textrm{Tr}\,e^{-\beta(H-\mu {N})}$ the partition function
is written as an imaginary time  integral $Z=\int D[\bar{\psi},\psi] e^{-\int_0^\beta d\tau \mathcal{L} }$, where the Lagrangian $\mathcal{L}= \sum_{\mathbf{k}}\bar{\psi}_{\mathbf{k}}(\tau)
\partial_{\tau}\psi_{\mathbf{k}}(\tau)+K$. Substituting for $K$ from Eq.\,(\ref{eqn:K}), we get
%
%\begin{equation*}
$\mathcal{L}=\sum_{\mathbf{k}}\bar{\psi}_{\mathbf{k}}(\tau)
(\partial_{\tau}-\mu+\epsilon_\mathbf{k}\uptau_{3})\psi_{\mathbf{k}}(\tau)
-\frac{V}{L^{d}}\sum_{\mathbf{q}}\bar{\Phi}(\mathbf{q},\tau)
{\Phi}(\mathbf{q},\tau).
$ %\end{equation*}
Next, Hubbard-Stratonovich fields, $\phi(\mathbf{q},\tau)$, are introduced to decouple the quartic $\bar{\Phi}(\mathbf{q},\tau)
{\Phi}(\mathbf{q},\tau)$ term. The resulting action is quadratic in the fermionic fields and can therefore be integrated out to give $Z=\int D[\phi^*]D[\phi]e^{-S_\textrm{eff}[\phi^*,\phi]}$, where
\begin{eqnarray}
S_\textrm{eff}[\phi^*,\phi]&=&-\textrm{Tr}\ln G^{-1} + \frac{L^d}{\beta V}\sum_q \left|\phi(q)\right|^2 \label{eqn:Seff}\\
G^{-1}(k,k')&=&\left(-i\epsilon_n - \mu +\epsilon_\mathbf{k}\uptau_3\right)\delta_{k,k'}\nonumber\\
&&-\frac{1}{\beta}\phi^*({k'-k})\uptau^+
-\frac{1}{\beta}\phi({k-k'})\uptau^-\label{eqn:Ginverse}
\end{eqnarray}
The  shorthand notations  $k\equiv(\mathbf{k},ik_n)$ and $k'-k = q \equiv (\mathbf{q},iq_m)$, where $k_n/q_m$ are the odd/even Matsubara frequencies, are used throughout. The Fourier transform  is defined as $\phi(x)={\beta^{-1}}\sum_m{L^{-d}}\sum_\mathbf{q} e^{-iq_m\tau + i\mathbf{q}\cdot \mathbf{r}} \phi(q).$ Since only real $\phi(x)$ is considered,  $\phi(q)=\phi^*(-q)$.

The mean-field solution corresponds to the saddle point $\phi(q)=\beta \delta_{m,0}\delta_{\mathbf{q},0}\phi_0$. The magnitude of $\phi_0$ is  obtained by minimizing the action  $\delta S_\textrm{eff}[\phi_0]/\delta \phi_0=0$. The saddle point condition at $T=0$ generalizes  Eq.\,(\ref{eqn:Vc}) to
\begin{equation}
\frac{1}{V}=\sum_\mathbf{k}\nolimits''\frac{1}{2E_\mathbf{k}}
\label{eqn:sce}
\end{equation} The mean-field scaling  $\phi_0\propto \sqrt{V-V_c}$ is recovered close to the QCP. Next, we investigate the stability of the saddle point solution by  analyzing  the Gaussian fluctuations around $\phi_0$. This is done by  expanding the action to quadratic order in  the deviation $\delta\phi(x)=\phi(x)-\phi_0$.

To this end, we first separate  the $\phi_0$ and the $\delta\phi$ contributions in $G$ in Eq.\,(\ref{eqn:Ginverse}) as $G^{-1}=G_0^{-1}(1-\Sigma G_0)$, where
\begin{eqnarray}
G_0(k,k')&=&-\frac{(i\epsilon_n+\mu)+\epsilon_\mathbf{k}\uptau_3-\phi_0\uptau_1}{(i\epsilon_n-\xi_\mathbf{k}^+)(i\epsilon_n-\xi^-_\mathbf{k})}\delta_{k,k'}\label{eqn:G0}\\
\Sigma(k,k')&=&\frac{1}{\beta}\delta\phi({k-k'})\uptau_1\label{eqn:Sigma}
\end{eqnarray}
The trace in Eq.\,(\ref{eqn:Seff}) is then expanded in the standard way using the formula   $\textrm{Tr} \ln G^{-1} = \textrm{Tr}\ln G_0^{-1}-\sum_n\frac{1}{n}\textrm{Tr}(G_0\Sigma)^n$. The  order $n=2$ terms are collected to derive the $|\delta\phi|^2$ corrections.  The first order correction, $n=1$, vanishes since the saddle-point is an extremum. We write the expansion to quadratic order of $S_\textrm{eff}$ as $S^{(2)}_\textrm{eff}=S_\textrm{eff}[\phi_0]+{L^d}\sum_q [V\Gamma(q)]^{-1} |\delta\phi(q)|^2 $, where $\Gamma^{-1}(q)=1-V\chi(q)$ is the generalization of Eq.\,(\ref{eqn:Uq}).

\begin{figure}[t]
\includegraphics[width=0.85\linewidth]{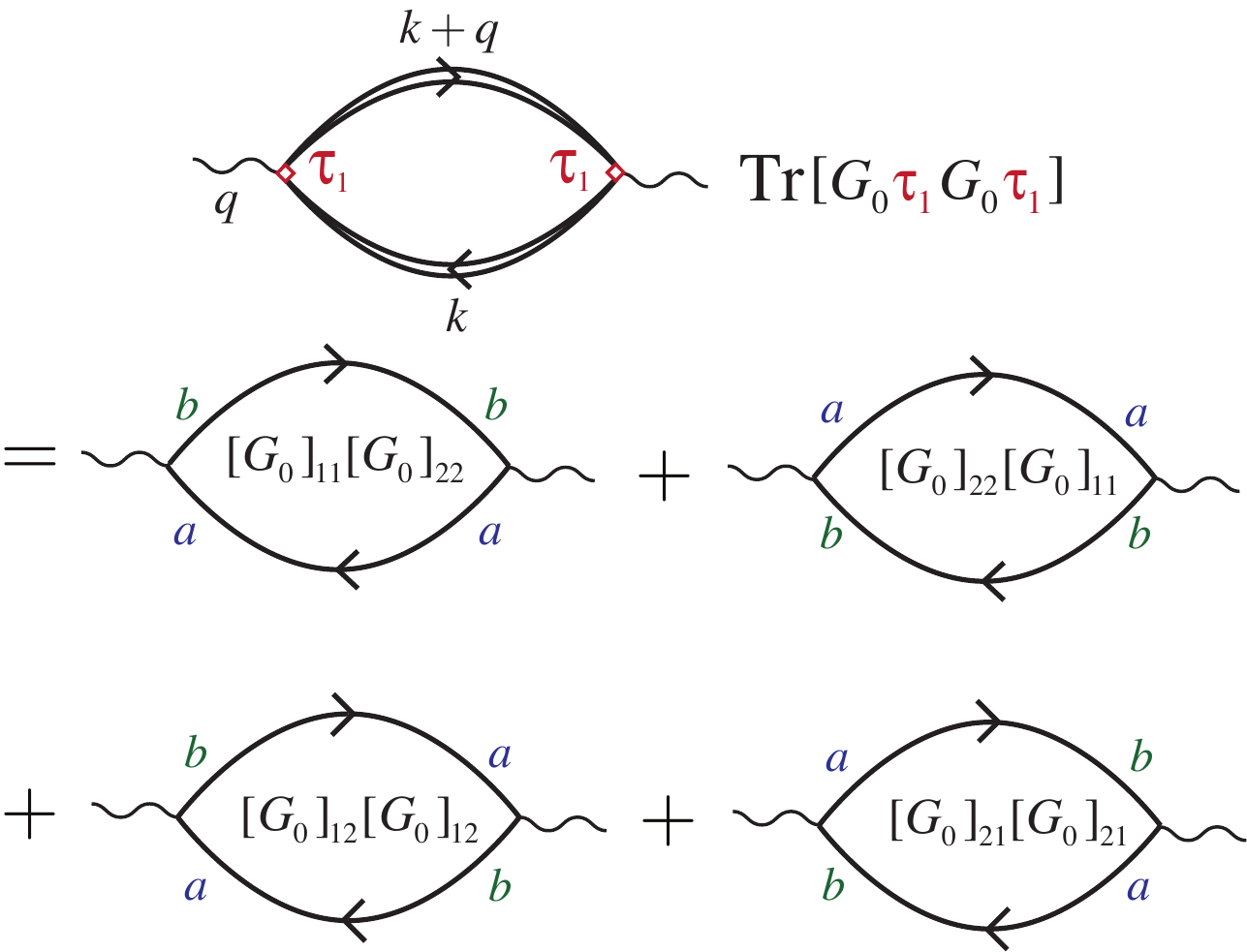}
\caption{The diagrammatic expansion of the trace in the definition of  $\chi(q)$  in Eq.\,(\ref{eqn:chi}) is shown. The new off-diagonal diagrams (last line) exist only in the XL phase, they vanish as $\phi_0^2$ close to the QCP. The interaction of the bosonic propagator, corresponding to the amplitude fluctuations of the order-parameter, with the gapless fermions at the chemical potential generates a bosonic mass $\sim \phi_0^2$. The sign of the mass is derived in the main text and is shown to be negative. }  \label{fig:chi}
\end{figure}

The  susceptibility $\chi(q)$ in the ordered phase equals
\begin{equation}
\chi(q)=-\frac{1}{2\beta L^d}\sum_k\textrm{Tr}\left[G_0(k+q)\uptau_1G_0(k)\uptau_1\right]
\label{eqn:chi}
\end{equation}
Note that the Green's function $G_0$ as defined in Eq.\,(\ref{eqn:G0}) is a $2\times 2$ matrix  written in the original FL basis $(b,a)$ and is therefore  not diagonal due to the hybridization of the bands in the XL phase.  Also note that the $\uptau_1$ in the trace originates from $\Sigma\sim \delta \phi \uptau_1$ (see Eq.\,(\ref{eqn:Sigma})). The diagrams corresponding to the various terms from the expansion of the trace are shown in Fig.\,\ref{fig:chi}. Compared to the interband FL susceptibility $\chi_0$ (see Eq.\,(\ref{eqn:chi0})), there are two new contributions in the XL phase that originate from the off-diagonal terms of $G_0$. They are shown  in the last line of Fig.\,\ref{fig:chi}, they vanish as $\sim \phi_0^2$.

 In the mean-field basis  $c_{\mathbf{k},\pm}$ defined in Eq.\,(\ref{eqn:uv}) (the XL basis) $\chi(q)$ separates into inter and intraband contributions, which we write as  $\chi(q)=\chi_\perp(q)+\chi_\parallel(q)$. After summing over the internal energy sums, we get
\begin{eqnarray}
\chi_\perp(q) &=&\frac{1}{2L^d} \sum_\mathbf{k}\nolimits'  \mathcal{F}_\perp(\mathbf{k,q}) \Bigl[\Pi_{+-}(\mathbf{k},q)+\Pi_{-+}(\mathbf{k},q) \Bigr]\hspace{0.6cm}\label{eqn:chi_perp}\\
\chi_\parallel(q)&=&\frac{1}{2L^d} \sum_\mathbf{k}\nolimits' \mathcal{F}_\parallel(\mathbf{k,q}) \Bigl[\Pi_{++}(\mathbf{k},q)+\Pi_{--} (\mathbf{k},q)\Bigr]\label{eqn:chi_parallel}
\end{eqnarray}
The form factors  $\mathcal{F}_\perp(\mathbf{k,q}) = (u_\mathbf{k}u_\mathbf{k+q}-v_\mathbf{k}v_\mathbf{k+q})^2$ and  $\mathcal{F}_\parallel(\mathbf{k,q})=(u_\mathbf{k}v_\mathbf{k+q}
+v_\mathbf{k}u_\mathbf{k+q})^2$, where  $u_\mathbf{k}/v_\mathbf{k}$ are defined in Eq.\,(\ref{eqn:uv}). The polarization functions
\begin{equation}
\Pi_{ss'}(\mathbf{k},q)=\frac{f(\xi_\mathbf{k+q}^s)-f(\xi^{s'}_\mathbf{k})}{
iq_m-(\xi_\mathbf{k+q}^{s}-\xi_\mathbf{k}^{s'})}
\end{equation}
where $s, s'=+/-$ are the upper/lower band indices.
Note that the polarization operator, $\Pi_{--}$, corresponds to the intraband particle-hole bubble  in the XL phase, i.e., the energies $\xi^\pm_\mathbf{k}$ correspond to the XL bands. (At $T=0$, the upper band is empty, hence $\Pi_{++}=0$.)

The hybridization of the bands in the XL phase couples the order-parameter to the gapless fermions at the Fermi surface. The  intraband polarization $\Pi_{--}$ represents this coupling, it vanishes as $\mathcal{F}_\parallel\sim \phi_0^2$ close to the QCP. Physically it generates an additional ``mass'' for the bosonic propagator. Depending on the sign of the mass, the condensate may or may not be stable. To determine the sign,  we expand both $\chi_\perp(q)$ and $\chi_\parallel(q)$ for small $q$ and take the static limit (mass term) of $\chi(q)$.

 Since $E_G>0$, the interband  terms $\Pi_{+-/-+}$ in $\chi_\perp$ have a regular expansion in $q$. After  analytical continuation, $iq_m=\omega+i0^+$, we get at $T=0$  to $\mathcal{O}(q^2,\phi_0^2)$
\begin{equation}
\chi_\perp(q)= \sum_\mathbf{k}\nolimits''\frac{1}{2E_\mathbf{k}} -\gamma_\perp(4 \phi_0^2-\omega^2+c {|\mathbf{q}|^2}/{2m})
\end{equation}
From the self-consistent equation derived in Eq.\,(\ref{eqn:sce}),  the first term equals $1/V$ at the saddle-point and cancels the constant in $\Gamma^{-1}(q)=1-V\chi(q)$. The remaining contributions of  $\mathcal{O}(\phi_0^2,q^2)$ to the bosonic propagator equal
\begin{equation}
[V \Gamma(q)]^{-1}=
{\gamma_\perp(4\phi_0^2-\omega^2+c|\mathbf{q}|^2/2m)-\chi_\parallel(q)}
\label{eqn:Gamma}
\end{equation}

 The occurrence of the pole $\omega^2=(2\phi_0)^2+c|\mathbf{q}|^2/2m$,  starting at $\omega=2\phi_0$,  is identified with the collective excitations of  the transverse field Ising model~\cite{deGennes_ising,brout_ising}. To elucidate this further, we rewrite the fermionic bilinears in the Hamiltonian $K$ in Eq.\,(\ref{eqn:K}) in terms of  pseudospin operators\cite{book_lipkin}
 $\sigma_i(\mathbf{k})=\psi^\dagger_\mathbf{k}\uptau_i\psi_\mathbf{k}$. They obey the usual $SU(2)$ commutation relations $[\sigma_i(\mathbf{k}),\sigma_j(\mathbf{k'})]
=2i\epsilon_{ijk}\sigma_k(\mathbf{k})\delta_{\mathbf{k,k'}}$.
The dynamics of the uniform state is described  by the $q=0$ term of $K$. The reduced Hamiltonian, $K_0$, equals
\begin{equation}
K_0=\sum_\mathbf{k}\epsilon_\mathbf{k}\sigma_3(\mathbf{k})
-\frac{V}{L^d}\sum_\mathbf{k,k'}\sigma^-(\mathbf{k})\sigma^+(\mathbf{k'})
\end{equation}
(The constant $\mu$ is suppressed here, it plays a crucial role only when the scattering at the Fermi surface involving $\mathbf{q}\neq 0$ are included.)
As $V$ increases, a transition occurs from the paramagnetic state in which all states point down (occupied valence band) to a correlated Ising ferromangnetic  state with spins pointing in the $x-y$ plane (hybridized states). Our choice of a real order-parameter,  $\langle \hat{\Phi}(0)\rangle = \langle \hat{\Phi}^\dagger(0)\rangle \propto \phi_0 $, breaks the symmetry  in the $x$ direction, i.e., $ \sum_\mathbf{k} \langle \sigma_1(\mathbf{k})\rangle\neq 0$ (see  Eq.\,(\ref{eqn:K_MF})). The key point is that this Ising symmetry is lost when the finite $\mathbf{q}\neq 0$ terms involving particle-hole scattering at the Fermi surface are included. The scattering is represented by  $\chi_\parallel(q)$, which  appears  as a self-energy correction to the bosonic propagator in  Eq.\,(\ref{eqn:Gamma}). We show next that the static limit  of $\chi_\parallel(q)\sim \phi_0^2$   is positive (implying a negative mass contribution to $\Gamma(q)$) and therefore has the potential  to destabilize the condensate. Note that at half-filling $\chi_\parallel=0$, hence the condensate is always stable.
 %[Remember th      bvec{k}=\sqrt{\epsilon_\mathbf{k}^2+\phi_0^2}$, and  the single prime denotes the upper cutoff specified in Eq.\,(\ref{eqn:cutoff})]%The ultraviolet cut-off is defined in Eq.\,(\ref{eqn:cutoff}).

 %
 First we show that the condensate is stable when    $\chi_\parallel$ is suppressed.
This requires  that the constants $\gamma_\perp$ and $c>0$ are positive in Eq.\,(\ref{eqn:Gamma}). The integral for $\gamma_\perp$ is ultraviolet  $(\Lambda)$ convergent in dimensions $d<6$  and hence we set $\Lambda\rightarrow \infty$ and restrict ourselves to $d<6$. In the limit $\phi_0,q=0$, only two parameters remain,  the dimension $d$ and the hole doping factor  $x=(1-E_G/2\epsilon_F)$.

Separating the factor $\nu(\epsilon)=\nu_d (\epsilon-E_G/2)^{d/2-1}$  where  $\nu_d=(m/2\pi)^{d/2}/\Gamma(d/2)$, corresponding to the density of states, the  constant $\gamma_\perp$ can be written as
 \begin{equation}
 \gamma_\perp = \sum_\mathbf{k}\nolimits''\frac{1}{(2\epsilon_\mathbf{k})^3}
=\frac{1}{8}\nu_d \epsilon_F^{d/2-3}R_\perp^{(d)}(x)
\label{eqn:gamma_perp}
\end{equation}
Restricting ourselves to $d=2,3$ and $4$, we get
\begin{equation}
\left\{R^{(2)}_\perp,R^{(3)}_\perp,R^{(4)}_\perp\right\}=\left\{\frac{1}{2},
\frac{\pi}{8}\Bigl(1+\frac{3x}{2}\Bigr),
\frac{1}{2}(1+x)\right\}
\end{equation}
Only the leading $\mathcal{O}(x)$ correction is shown in $d=3$.  We find that the constant $c=2E_G/(4-d)+\mathcal{O}(x)$, which is ultraviolet convergent only for $d<4$. (The derivation of $c$ is not shown here as it is not crucial to the discussion.) We therefore restrict our analysis to below $d=4$. %For the constant $c$, we get   $c=[{2E_G}/({4-d})](1 + \mathcal{O}(x))$ for $d<4$.
%
%Hence,  the FL to XL transition belongs to the universality class of the transverse field Ising model when only the interband terms are retained.
%

 Finally, the effect of the self-energy correction $\chi_\parallel$  in Eq.\,(\ref{eqn:Gamma}) is derived.
From Eq.\,(\ref{eqn:uv}) it  follows that the form-factor  $\mathcal{F}_\parallel\sim (u_\mathbf{k}v_\mathbf{k})^2\sim (\phi_0/2E_\mathbf{k})^2$.
Hence to $\phi_0^2$ order, it is sufficient to set $\phi_0=0$ in the polarization function: $\chi_\parallel (q)= {\phi_0^2}\sum_\mathbf{k}\nolimits''\Pi_0(q)/2\epsilon_\mathbf{k}^2$, where $\Pi_0(q)=\Pi_{--}(q)|_{\phi_0=0}$  is the standard Lindhard-type function.  It has the well known FL singularities
\begin{equation}
\Pi^0(q) \approx -\frac{\partial f(\xi_\mathbf{k}^-)}{\partial \xi^-_\mathbf{k}}\times\left(\frac{-\mathbf{q}\cdot \nabla_\mathbf{k}\xi^-_\mathbf{k}}{\omega -\mathbf{q}\cdot \nabla_\mathbf{k}\xi^-_\mathbf{k}}\right)
\label{eqn:Pi0}
\end{equation}
%
%$\Pi^0(q) \approx \frac{\partial f(\xi_\mathbf{k}^-)}{\partial \xi^-_\mathbf{k}}\times\left(\frac{-\mathbf{q}\cdot \nabla_\mathbf{k}\xi^-_\mathbf{k}}{\omega -\mathbf{q}\cdot \nabla_\mathbf{k}\xi^-_\mathbf{k}}\right)$%
with different limiting values~\cite{bluebook} when $\omega, |\mathbf{q}|\rightarrow 0$: the dynamic limit $\Pi_0^\omega=\lim_{\omega\rightarrow 0}\Pi_0(\mathbf{q}=0,\omega)= 0$, and the static limit $\Pi_0^q=\lim_{|\mathbf{q}|\rightarrow 0}\Pi_0(\mathbf{q},\omega=0)=- \partial f(\xi_\mathbf{k}^-)/\partial \xi^-_\mathbf{k}$.  Importantly, this FL singularity induces a singularity in $\chi_\parallel(q)$ which in turn induces a singularity in   $\Gamma(q)$.
Keeping the same  notation for the static and dynamic limits, we get $\chi_\parallel^q = - \gamma_\parallel(2\phi_0)^2$, where
\begin{equation}
\gamma_\parallel =    \frac{1}{8}{\nu_d}\epsilon_F^{d/2-3}R_\parallel^{(d)}(x)
\label{eqn:gamma_parallel}
\end{equation}
The function $R^{(d)}_\parallel(x)= - x^{d/2-1}$ derives its form from the density of states at the chemical potential.

We now combine the two contributions $\gamma_\perp + \gamma_\parallel$  to determine the sign of the mass term.
To this end, we write the  static limit $\Gamma^q$ in Eq.\,(\ref{eqn:Gamma}) as
\begin{equation}
[V\Gamma^q]^{-1}\equiv \gamma(2\phi_0)^2=
\frac{1}{8}{\nu_d}\epsilon_F^{d/2-3}R^{(d)}(x)(2\phi_0)^2
\label{eqn:Gamma_q}
 \end{equation}
From  Eqs.\,(\ref{eqn:gamma_perp}) and (\ref{eqn:gamma_parallel}) we obtain the following expressions for the leading $x$ dependence for $R^{(d)}=R_\perp^{(d)}+R_\parallel^{(d)}$ in dimensions $d=2,3$ and $4$ %
\begin{equation}
\left\{R^{(2)},R^{(3)},R^{(4)}\right\}=\left\{-\frac{1}{2},
\left(\frac{\pi}{8}-\sqrt{x}\right),\frac{1}{2}(1-x)\right\}
\label{eqn:gamma}
\end{equation}
Note that the filling factor $x<1$ away from half-filling. Hence for all $d<4$ the function $R^{(d)}(x)$ becomes negative (and correspondingly $\gamma$ becomes negative) beyond some  critical  doping $x_c$.  In particular,  $x_c=0$ in $d=2$, and $x_c = (\pi/8)^2\approx 0.15$ in $d=3$   (the exact value is slightly higher at $\approx 0.2$ corresponding  to a  hole Fermi energy $\sim E_G/8$).  A negative $\gamma$ in Eq.~(\ref{eqn:Gamma_q}) implies an instability of the XL phase (negative mass).

This is our main result in this section, namely, the uniform XL obtained self-consistently at the mean-field level is mostly unstable in $d<4$ (small doping) and is always unstable in $d=2$.  %this is our main result in this section. %At $d=4$, and presumably beyond, it never becomes negative.% It may be that a finite $Q\sim \nu_d \phi_0^2/E_G$ state may be stable. This question is not addressed in this paper.

\section{XL phase with broken \texorpdfstring{$\mathcal{T}$}{mathT} symmetry}

In the last section, we demonstrated that the uniform XL phase with $s$-wave pairing in a two-band system with quadratic dispersions is unstable in $d=2$. We extend this analysis  to study the stability of the XL phase in an interacting system in which $\mathcal{T}$ symmetry is spontaneously broken in the groundstate. To this end, we consider the  extended Hubbard model of spinless electrons  on a honeycomb lattice with  next-nearest-neighbor ($nnn$) repulsion. The Hamiltonian is defined as
\begin{equation}
H=-t\sum_{nn}(A^\dagger_iB_j+B^\dagger_iA_j)
+V\sum_{nnn}(n_i^An_j^A+n_i^Bn_j^B)
\label{eqn:H_realspace_graphene}
\end{equation}
The operators $A^\dagger_i/A_i$ and $B^\dagger_i/B_i$ are the on-site electron creation/annihilation operators on the respective  sub-lattices  and $n_i^{A/B}$ are the number operators. The spin degrees of freedom are suppressed  to avoid any spin-related $\mathcal{T}$ symmetry breaking effects~\cite{FM_volkov}.

\begin{figure}[t]
\includegraphics[width=0.75\linewidth]{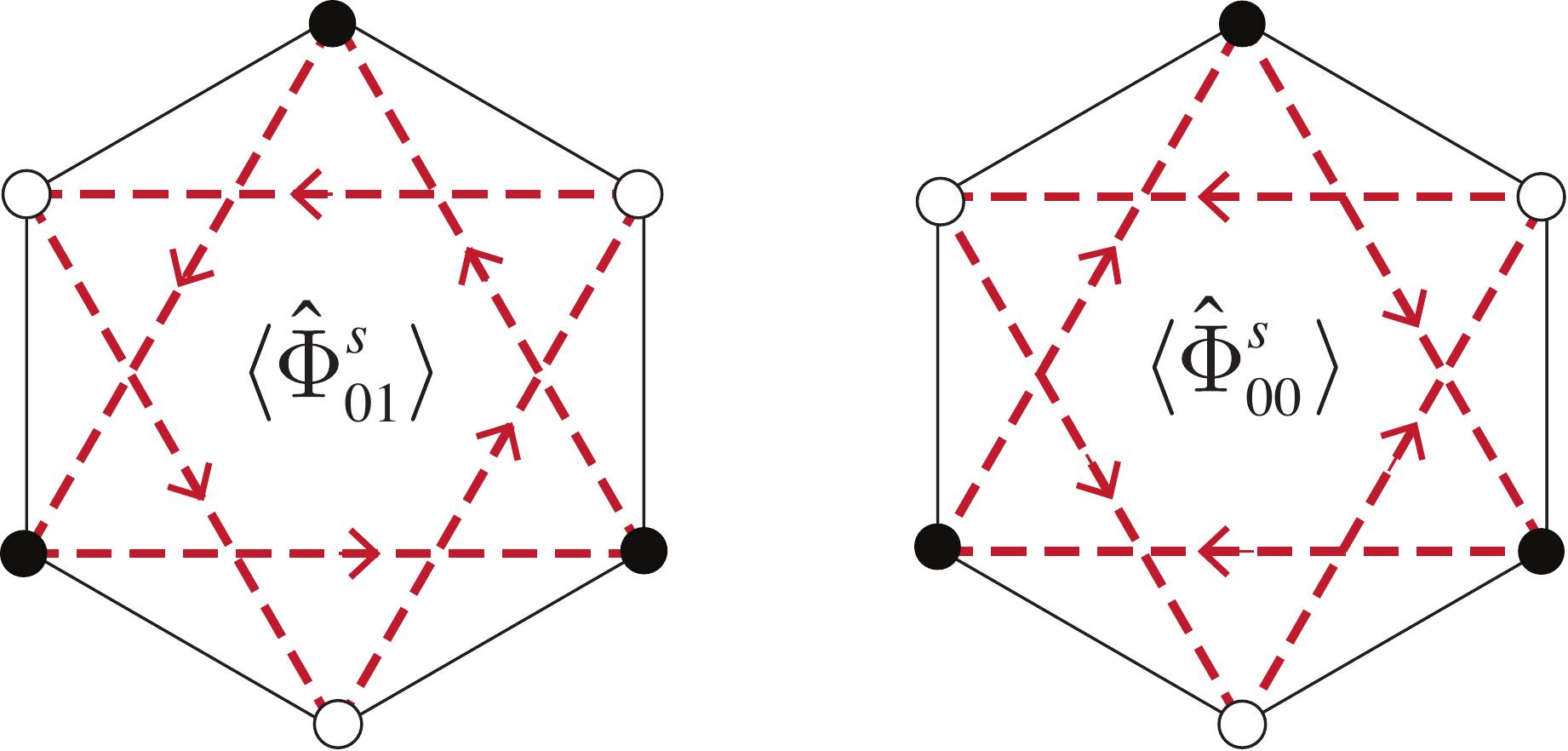}
\caption{The ordered states  can be pictured as two separate current carrying  loops running along the sides  of the two triangular sub-lattices composed of $A$ (white vertices) and $B$ (black vertices) sites that form the honeycomb lattice. (The atoms at $A$ and $B$ are assumed identical.) The  current loops break the $\mathcal{T}$ symmetry~\cite{haldane_graphene_TRV}. Since the total moment in a unit cell is zero, they generate translationally invariant patterns when extended over the whole lattice.  The pattern generated by $\langle \hat{\Phi}_{0 1}^{s\dagger}(\mathbf{q}=0)\rangle$ has $C_6$ symmetry, while $\langle \hat{\Phi}_{0 0}^{s\dagger}(\mathbf{q}=0)\rangle$ has only a reduced $C_{3v}$ symmetry that breaks inversion symmetry.}  \label{fig:loops}
\end{figure}

The non-interacting spectrum with nearest-neighbor ($nn$) tunneling is a semi-metal with two inequivalent degeneracy points,  $\pm \mathbf{k}_D$, called Dirac points,  located at the corners of the Brillouin Zone. %, where $\mathbf{k}_D=(2\pi/3\sqrt{3}a, 2\pi/3a)$.
%The gap vanishes vanishes at the Dirac points~\cite{wallace_1947}, i.e.,  $\epsilon_{\mathbf{k}_D}=0$.
At the mean-field level, beyond a critical interaction strength $V_c$ the interaction lifts the degeneracy at the Dirac points and opens a gap to stabilize a Topological Mott Insulator~\cite{Zhang_TMI} (TMI) phase at \textit{half-filling}.   Unlike the conventional Mott phase, the TMI breaks both chirality ($\mathcal{C}$) and $\mathcal{T}$ symmetries but is invariant under the combined $\mathcal{CT}$ transformation. It is therefore a type-II state according to the classification in Ref.\,\onlinecite{Fradkin_sun_TRV}. (See Fig.\,\ref{fig:loops} for a physical description of these states.) The properties of the TMI phase are  insensitive to the $nn$ repulsion $U$ when $U/V_c\ll 1$. We therefore  neglect $U$ in our model. (Further details about the interplay of $U$ and $V$ can be found in Ref.\,\onlinecite{Zhang_TMI}.)

The methods developed in the last section are applied here to study the stability of the broken $\mathcal{T}$  phase away from half-filling. We first write the Hamiltonian (\ref{eqn:H_realspace_graphene}) in $\textbf{k}$-space  (we set $t=1$ and use $N$ for the number of sites)
\begin{eqnarray}
H&=&-\sum_\mathbf{k}\epsilon_\mathbf{k}e^{-i\theta_\mathbf{k}}
A_\mathbf{k}^\dagger B_\mathbf{k}-\sum_\mathbf{k}\epsilon_\mathbf{k}
e^{i\theta_\mathbf{k}}B_\mathbf{k}^\dagger A_\mathbf{k}\nonumber\\
&&-\frac{V}{N}\sum_\mathbf{k,p;q} g(\mathbf{k-p})A_{\mathbf{k}+\frac{\mathbf{q}}{2}}^\dagger A_{\mathbf{k}-\frac{\mathbf{q}}{2}}A_{\mathbf{p}-\frac{\mathbf{q}}{2}}^{\dagger}
A_{\mathbf{p}+\frac{\mathbf{q}}{2}}\nonumber\\
&&-\frac{V}{N}\sum_{\mathbf{k,p;q}}g(\mathbf{k-p})
B_{\mathbf{k}+\frac{\mathbf{q}}{2}}^{\dagger}
B_{\mathbf{k}-\frac{\mathbf{q}}{2}}
B_{\mathbf{p}-\frac{\mathbf{q}}{2}}^{\dagger}
B_{\mathbf{p}+\frac{\mathbf{q}}{2}}
\label{eqn:Hk}
\end{eqnarray}
The real part of the kinetic term $\epsilon_{\mathbf{k}}e^{i\theta_{\mathbf{k}}}
=\sum_{l}\exp(i\mathbf{k}\cdot\mathbf{d}_{l})$ gives the energy $\epsilon_\mathbf{k}=\sqrt{3+2\sum_l \cos(\mathbf{k}\cdot\mathbf{t}_l)}$,  it vanishes  at the Dirac points, i.e., $\epsilon_{\pm \mathbf{k}_D}=0$.  The $\mathbf{d}_l$ vectors connect the nearest neighbor atoms and $\mathbf{t}_l$'s are the basis vectors of the hexagonal Bravais lattice.  ($l=1,2$ and $3$). In the following, we transform to the basis in which the kinetic term is diagonal, namely,
\begin{equation}
\psi_\mathbf{k}=\left(\begin{array}{c}
b_\mathbf{k}\\ a_\mathbf{k}
\end{array}\right) =\frac{1}{\sqrt{2}}\left(\begin{array}{cc}
-e^{\frac{i}{2}\theta_\mathbf{k}} & e^{-\frac{i}{2}\theta_\mathbf{k}}\\
e^{\frac{i}{2}\theta_\mathbf{k}} & e^{-\frac{i}{2}\theta_\mathbf{k}}
\end{array}\right)\left(\begin{array}{c}
B_\mathbf{k}\\ A_\mathbf{k}
\end{array}\right)
\label{eqn:spinor}
\end{equation}

The various  interaction induced symmetry breaking possibilities consistent with the lattice symmetries can be gleaned by expressing $g(\mathbf{k-p})=\sum_l
\cos[(\mathbf{k-p})\cdot\mathbf{t}_{l}]$ in terms of the distinct irreducible representations (irreps) of the underlying lattice.   One possible decomposition involving separable irreps of the planar $C_{6v}$ symmetry group of the honeycomb lattice is shown below
\begin{equation}
g(\mathbf{k-p})=\sum_{\nu=0}^{2}
\mathcal{A}_{\nu}^{s*}(\mathbf{k})\mathcal{A}_{\nu}^{s}(\mathbf{p})
+\mathcal{A}_{\nu}^{c*}(\mathbf{k})\mathcal{A}_{\nu}^{c}(\mathbf{p})
\end{equation}
Here, $\mathcal{A}^{s/c}_0$ belong to the  one-dimensional representations $B_1/A_1$, while  $(\mathcal{A}^{s/c}_1,\mathcal{A}^{s/c}_2)$ form the basis for  the two-dimensional representations $E_1/E_2$. (See, e.g., Ref.\,\onlinecite{tinkham_book} for the notation used here.)
Explicitly, the functions  $\mathcal{A}_{\nu}^{s}(\mathbf{k})
 =\frac{1}{\sqrt{3}}[\sin(\mathbf{k}\cdot\mathbf{t}_{1})
 +e^{i\frac{2\nu\pi}{3}}\sin(\mathbf{k}\cdot\mathbf{t}_{2})
 +e^{i\frac{4\nu\pi}{3}}\sin(\mathbf{k}\cdot\mathbf{t}_{3})]$, and the  $\mathcal{A}_{\nu}^{c}(\mathbf{k})$ functions have  sines  replaced by cosines.

We first note that in order to have the closed loop configurations  shown in Fig\,\ref{fig:loops}, it is necessary that the order-parameter has a minimum of $C_3$ symmetry. The two-dimensional irreps $E_1$ and $E_2$\ have characters  -1 under $C_3$ and therefore can not  form closed loops.  Furthermore, since  under a time-reversal operation an arbitrary function $\mathcal{A}(\mathbf{k})$ transforms as $\mathcal{T} \mathcal{A}(\mathbf{k}) \mathcal{T}^{-1}=\mathcal{A}^*(-\mathbf{k})$, only the $B_1$ representation, or $\mathcal{A}^s_0$, breaks the $\mathcal{T}$ symmetry. ($\mathcal{A}_0^c$ does not break $\mathcal{T}$ symmetry.) $\mathcal{A}^s_0$ has a reduced  $C_6$ symmetry  and hence breaks the $C_{6v}$ symmetry  to $C_6$.%Note that besides $\mathcal{T}$ symmetry, $\mathcal{A}^s_0$ also breaks the $C_{6v}$ symmetry down to $C_6$.  %Note that $T^s_0(\mathbf{k})$ is real in momentum-space and hence is complex in real-space.

In the ladder approximation, the interactions in the $\mathcal{A}^s_0$ channel  diverge near the  QCP and hence the other channels can be neglected. This simplifies the interaction part of the Hamiltonian (\ref{eqn:Hk}) to just two terms
\begin{equation}
H_{int} = -\frac{V}{2N}\sum_\mathbf{q}\hat{\Phi}_{0 1}^{s\dagger}(\mathbf{q})\hat{\Phi}_{0 1}^{s}(\mathbf{q})+\hat{\Phi}_{0 0}^{s\dagger}(\mathbf{q})\hat{\Phi}_{0 0}^{s}(\mathbf{q})
\end{equation}
The bilinear operators are obtained by combining the $A_\mathbf{k}$ and $B_\mathbf{k}$ operators as $\psi_\mathbf{k}$ (see Eq.\,(\ref{eqn:spinor}))
\begin{equation}
\hat{\Phi}_{0 j}^{s}(\mathbf{q})=\sum_{\mathbf{k}}S_\mathbf{k}\psi_{\mathbf{k}+\frac{\mathbf{q}}{2}}^{\dagger}
e^{\frac{i}{2}\theta_{\mathbf{k}+\frac{\mathbf{q}}{2}}\uptau_1}
\uptau_{j} e^{-\frac{i}{2}\theta_{\mathbf{k}-\frac{\mathbf{q}}{2}}\uptau_1}
\psi_{\mathbf{k}-\frac{\mathbf{q}}{2}}
\label{eqn:Phi}
\end{equation}
For notational simplicity, we define $S_\mathbf{k}\equiv\mathcal{A}^s_0(\mathbf{k})$.
Note that our choice of the basis in Eq.\,(\ref{eqn:spinor})  introduces phases in the interaction. The indices $j=0,1$ label the Pauli matrices $\uptau_0$ (identity) and $\uptau_1$. Physically they correspond to adding or subtracting the two loop currents on each sublattice as described in  Fig.\,\ref{fig:loops}.
Mean-field analysis~\cite{Zhang_TMI} of the two patterns favors the condensation of $\langle \hat{\Phi}_{0 1}^{s\dagger}(\mathbf{q}=0)\rangle$.  We therefore neglect $\hat{\Phi}_{0 0}$  and arrive at our minimal model
\begin{equation}
K_T =\sum_{\mathbf{k}}\psi_{\mathbf{k}}^{\dagger}[\epsilon_{\mathbf{k}}\uptau_3-\mu]
\psi_{\mathbf{k}} -\frac{V}{2N}\sum_\mathbf{q}\hat{\Phi}_{0 1}^{s\dagger}(\mathbf{q})\hat{\Phi}_{0 1}^{s}(\mathbf{q})
\label{eqn:KT}
\end{equation}

The Hamiltonian $K_T$ is a generalization of $K$ derived in Eq.\,(\ref{eqn:K}) and
the  mean-field analysis and the fluctuation calculations follow exactly as detailed in the last section.  %(The mean-field results at half-filling have been derived in \cite{Zhang_TMI}.)
To avoid repetition we present only the main steps below. %[Similar conclusions as (i) and (ii) below have been obtained in the half-filled case in Ref.\,\cite{Zhang_TMI}.]

\subsection{Mean-field theory}

%The mean-field energies and the coefficients of the wavefunctions  are derived below.

We note that since  ${\hat{\Phi}_{0 1}^{s\dagger}}(\mathbf{q})={\hat{\Phi}_{0 1}^{s}}(-\mathbf{q})$, the order-parameter $\phi_T=\frac{V}{N}\langle \hat{\Phi}_{0 1}^{s}(0)\rangle$ is real. Hence, the mean-field Hamiltonian reads
\begin{eqnarray}
K_{T,\text{MF}}=\sum_{\mathbf{k}}\psi_{\mathbf{k}}^{\dagger}\left[\epsilon_{\mathbf{k}}\uptau_{3}-\mu-\phi_{T}S_{\mathbf{k}}\uptau_{1}\right]\psi_{\mathbf{k}}+\frac{N}{2V}\phi_{T}^{2}\hspace{0.5cm}
\label{eqn:KT_MF}
\end{eqnarray}
It follows  that the  energies $\xi^\pm_\mathbf{k}=\pm E_\mathbf{k}-\mu$, where $E_\mathbf{k}=\sqrt{\epsilon_\mathbf{k}^2+\phi_T^2S_\mathbf{k}^2}$. Thus, a gap $\sim \phi_T |S_{\mathbf{k}_D}|$ opens at  $\pm\mathbf{k}_D$. However, since  $S_\mathbf{k}$ is an odd function, the sign of the gaps are in  opposite directions at the two Dirac points. This is the origin of the anomalous quantum Hall effect as described by Haldane~\cite{haldane_graphene_TRV}. A non-zero $\phi_T$ breaks  $\mathcal{T}$ symmetry but does not break inversion symmetry~\cite{IT_graphene}.

 Minimizing the action we obtain the self-consistent equation at $T=0$ as $1/V=\frac{1}{N}\sum_{BZ}\nolimits' S_\mathbf{k}^2/E_\mathbf{k}$. As before, we consider hole doping, i.e., $\mu<0$. At $T=0$, all states $|\mathbf{k}| > k_F$ measured from $\pm \mathbf{k}_D$ are filled, leaving small hole pockets at the Dirac points. The lower cutoff $k_F$ is denoted by the  prime on the summation. Since it is a  convergent integral no upper-cutoff is required and  the summation extends over the whole Brillouin Zone ($BZ$).

Finally, as before we write the eigenfunctions in the form  $c_{\mathbf{k}+}= u_\mathbf{k} b_\mathbf{k}-v_\mathbf{k}a_\mathbf{k} $ and $c_{\mathbf{k}-}=u_\mathbf{k}a_\mathbf{k} +  v_\mathbf{k} b_\mathbf{k}$, corresponding to $\pm E_\mathbf{k}$. After minimizing the  free energy, we get for $u_\mathbf{k}^2$ and $v_\mathbf{k}^2$ the same relations as  in Eq.\,(\ref{eqn:uv}) with an additional phase for $v_k$ given as ($\phi_T>0$ is assumed)
\begin{equation}
v_\mathbf{k}=\sqrt{v_\mathbf{k}^2}\,\frac{S_\mathbf{k}}{|S_\mathbf{k}|}=\sqrt{v_\mathbf{k}^2}\ \textrm{sign}(S_\mathbf{k})
\label{eqn:vk_T}
\end{equation}
This additional phase does not appear in the calculation of the static susceptibility as shown below.

\subsection{Order-parameter fluctuations}
Since only the stability of the condensate is under question, we only examine the static limit of the susceptibilities below. We follow the same steps as in the previous section. First, the matrices in Eqs.\,(\ref{eqn:G0}) and (\ref{eqn:Sigma}) are modified for the honeycomb lattice as:
\begin{eqnarray}
G_{0}(k,k')&=&-\frac{(i\epsilon_n+\mu)\uptau_0+\epsilon_\mathbf{k}\uptau_3-\phi_T S_\mathbf{k}\uptau_1}
{(i\epsilon_{n}-\xi_\mathbf{k}^+)(i\epsilon_n-\xi_\mathbf{k}^-)}\delta_{k,k'}\hspace{0.5cm}\\
\Sigma(k,k')&=&\frac{1}{\beta}\delta\phi(k-k')
S_{\mathbf{k}_+}e^{\frac{i}{2}(\theta_\mathbf{k}-\theta_\mathbf{k'})\uptau_1}\uptau_1
\end{eqnarray}
The momentum $\mathbf{k}_+=(\mathbf{k+k'})/{2}$. Secondly, the form factors in Eqs.\,(\ref{eqn:chi_perp}) and (\ref{eqn:chi_parallel}) are modified to include the structure factor $S_\mathbf{k}$  as
$\mathcal{F}_\perp(\mathbf{k},\mathbf{q}=0)= S_\mathbf{k}^2 (u_\mathbf{k}^2-v_\mathbf{k}^2)$ and $\mathcal{F}_\parallel(\mathbf{k},\mathbf{q}=0)=4 S_\mathbf{k}^2 u_\mathbf{k}^2v_\mathbf{k}^2$. As noted  below Eq.\,(\ref{eqn:vk_T}), only the squares of $u_\mathbf{k}$ and $v_\mathbf{k}$ appear in these expressions.
%
%The expressions for $u_\mathbf{k}/v_\mathbf{k}$  are the same as in (\ref{eqn:uv}).

Finally, to $\mathcal{O}(\phi_T^2)$ we get $[V\Gamma^q]^{-1}=\gamma \phi_T^2 $ for the static limit of the interaction amplitude [equivalent to Eq.\,(\ref{eqn:Gamma_q})]
\begin{equation}
\gamma=\frac{1}{N}\sum_\mathbf{k}\nolimits' \frac{S_\mathbf{k}^4}{\epsilon_\mathbf{k}^2}\left[\frac{f(\xi_\mathbf{k}^-)}{\epsilon_\mathbf{k}} + \frac{\partial f(\xi_\mathbf{k}^-)}{\partial \xi_\mathbf{k}^-} \right]
\label{eqn:gamma_graphene}
\end{equation}
It includes contributions from the inter and intraband  susceptibility, i.e.,  $\gamma=\gamma_\perp +\gamma_\parallel$, as explained in Eqs.\,(\ref{eqn:gamma_perp}) and (\ref{eqn:gamma_parallel}). Since $S_\mathbf{k}^4$ is a sharply peaked function around $\mathbf{k}_D$, the integrals can be evaluated by linearizing the spectrum around $\mathbf{k}_D$ as $\epsilon_\mathbf{k}\sim \alpha |\mathbf{k}-\mathbf{k}_D|$, where $\alpha=3/2$ (in units of lattice spacing).  The lower cut-off is from $k_F$ and hence there is no divergence away from half-filling and an expansion in $\phi_T^2$ is possible (unlike at half-filling\,\cite{Zhang_TMI}). %Again we notice that since the derivative of the Fermi function is negative, the intraband contribution  acts to destabilize the ordered XL phase.

After linearizing, we get
%
%\begin{equation}
$\gamma_\parallel = - 2\nu_0 S_{k_F}^4/\epsilon_F  $. The negative sign originates from the derivative of the Fermi function.
%\end{equation}
%$S_F$ corresponds to the value of $S_\mathbf{k}$ at $|k|=k_F$.
The density of states around each $\mathbf{k}_D$ equals $\nu_0=1/\sqrt{3}\pi$ and the factor $2$ accounts for the contributions from $\pm \mathbf{k}_D$. The interband contribution is easily shown to satisfy $\gamma_\perp< |\gamma_\parallel|$, implying that  $\gamma <0$ in Eq.\,(\ref{eqn:gamma_graphene}). Hence, we conclude that the uniform $\mathcal{T}$ broken state on the honeycomb lattice, which is stable at half-filling \cite{Zhang_TMI}, is unstable to infinitesimal doping.

\section{Discussion}
We have shown that the mean-field solution for the  interband particle-hole condensate with a sharp Fermi surface, which we call an excitonic liquid (XL) in this paper, is unstable in the presence of the  gapless  fermions at the Fermi surface. The origin of the instability is closely related to the singularity of the FL polarization function in Eq.\,(\ref{eqn:Pi0}).
We demonstrate this destabilization in two models, both of which stabilizes an uniform XL phase at the mean-field level. Our results therefore suggest that  a uniform condensate of virtual excitons, with or without spontaneous time reversal symmetry breaking, is an unstable phase at $T=0$. We arrive at this conclusion by analyzing the static limit of the effective interaction in the particle-hole channel and showing it to be negative. It follows that  a Ginzburg-Landau type description of the ordered phase~\cite{Fradkin_sun_TRV} is in general not possible.

Finally, a few remarks about  the relevance of the higher order terms in the expansion of the action $S_\textrm{eff}$ in Eq.\,(\ref{eqn:Seff}). In general, the bosonic action can be expanded as
\begin{equation}
S_\textrm{eff}=S_\textrm{eff}^{(2)} +\sum_{n=2}^\infty\int(d\omega d^2\textbf{q})^{2n-1}b_{2n}(\delta\phi(q))^{2n}
\end{equation}
The coefficients $b_{2n}$  were calculated in  Ref.~\onlinecite{abanov_chubukov} and shown to contain universal singular contributions that makes the coefficients anomalously large in the dynamic limit $\omega\rightarrow 0$ and $q=0$, leading to the general conclusion that the Hertz theory is incomplete. (The corresponding dynamical corrections in $S_\textrm{eff}^{(2)}$ is  the familiar Landau damping term.) In the opposite limit, i.e., the static limit, no such anomalous contributions exists and the Hertz assumption that the vertices are local is restored, allowing for a controlled expansion in powers of $\delta\phi$. To establish the stability of the mean-field groundstate, it is necessary that the static susceptibilites for all possible perturbations of the groundstate are positive. For  the order parameter fluctuations, this translates to the sign of the coefficient of the $(\delta\phi)^2$ term in a finite field $\phi_0$. Close to the transition, we calculate the contributions to order  $\phi_0^2$, which are shown in Fig.~\ref{fig:chi}. Note that there are no vertex corrections in this order.

Our results suggest that XL condensates with a sharp Fermi surface tends to be unstable in $d=2$. Of course, any condensate  can be stabilized if a gap opens, however, the mechanism to open a gap in the XL condensate  is unclear at the moment\,\cite{varma_1999}.
%%We believe that the methods developed in this paper will be useful to arrive at a more general proof.
%
Another possible cure for this instability might be to assume a non-uniform ($q\neq 0$) mean-field state similar to the spin-bag models proposed in the case of doped anti-ferromagnetism~\cite{AFM_Singh_1,AFM_Singh_2}. This possibility is not analyzed in this paper.
%
%The approach given in this paper allows us  to systematically formulate and study the current-loop fluctuations of the orbital currents in  the 3-band CuO system\,\cite{Varma_1997}. Work in this direction is in progress.

\section{Acknowlegements}
We acknowledge insightful conversations with  J.\ Birman (CCNY),  R.\ Matheus (IFT), A.\ R.\ Rocha (IFT)  and C.\ M.\ Varma (UCR). We would like to thank the CMB and the  Physics Departments at Tulane University for their kind hospitality where part of this work was done. Partial support was provided by PSC-CUNY Award 41.
%\bibliographystyle{alpha}
%\bibliography{TRV}

\begin{thebibliography}{38}%
\makeatletter
\providecommand \@ifxundefined [1]{%
 \@ifx{#1\undefined}
}%
\providecommand \@ifnum [1]{%
 \ifnum #1\expandafter \@firstoftwo
 \else \expandafter \@secondoftwo
 \fi
}%
\providecommand \@ifx [1]{%
 \ifx #1\expandafter \@firstoftwo
 \else \expandafter \@secondoftwo
 \fi
}%
\providecommand \natexlab [1]{#1}%
\providecommand \enquote  [1]{``#1''}%
\providecommand \bibnamefont  [1]{#1}%
\providecommand \bibfnamefont [1]{#1}%
\providecommand \citenamefont [1]{#1}%
\providecommand \href@noop [0]{\@secondoftwo}%
\providecommand \href [0]{\begingroup \@sanitize@url \@href}%
\providecommand \@href[1]{\@@startlink{#1}\@@href}%
\providecommand \@@href[1]{\endgroup#1\@@endlink}%
\providecommand \@sanitize@url [0]{\catcode `\\12\catcode `\$12\catcode
  `\&12\catcode `\#12\catcode `\^12\catcode `\_12\catcode `\%12\relax}%
\providecommand \@@startlink[1]{}%
\providecommand \@@endlink[0]{}%
\providecommand \url  [0]{\begingroup\@sanitize@url \@url }%
\providecommand \@url [1]{\endgroup\@href {#1}{\urlprefix }}%
\providecommand \urlprefix  [0]{URL }%
\providecommand \Eprint [0]{\href }%
\providecommand \doibase [0]{http://dx.doi.org/}%
\providecommand \selectlanguage [0]{\@gobble}%
\providecommand \bibinfo  [0]{\@secondoftwo}%
\providecommand \bibfield  [0]{\@secondoftwo}%
\providecommand \translation [1]{[#1]}%
\providecommand \BibitemOpen [0]{}%
\providecommand \bibitemStop [0]{}%
\providecommand \bibitemNoStop [0]{.\EOS\space}%
\providecommand \EOS [0]{\spacefactor3000\relax}%
\providecommand \BibitemShut  [1]{\csname bibitem#1\endcsname}%
\let\auto@bib@innerbib\@empty
%</preamble>
\bibitem [{\citenamefont {Mott}(1961)}]{Mott_Transition}%
  \BibitemOpen
  \bibfield  {author} {\bibinfo {author} {\bibfnamefont {N.~F.}\ \bibnamefont
  {Mott}},\ }\href@noop {} {\bibfield  {journal} {\bibinfo  {journal} {Philos.
  Mag.}\ }\textbf {\bibinfo {volume} {6}},\ \bibinfo {pages} {287} (\bibinfo
  {year} {1961})}\BibitemShut {NoStop}%
\bibitem [{\citenamefont {Knox}(1963)}]{Knox}%
  \BibitemOpen
  \bibfield  {author} {\bibinfo {author} {\bibfnamefont {R.~S.}\ \bibnamefont
  {Knox}},\ }\enquote {\bibinfo {title} {Theory of excitons},}\ in\ \href@noop
  {} {\emph {\bibinfo {booktitle} {Solid State Physics Suppl.}}},\
  Vol.~\bibinfo {volume} {5},\ \bibinfo {editor} {edited by\ \bibinfo {editor}
  {\bibfnamefont {F.}~\bibnamefont {Seitz}}\ and\ \bibinfo {editor}
  {\bibfnamefont {D.}~\bibnamefont {Turnbull}}}\ (\bibinfo  {publisher}
  {Academic Press, Inc., New York},\ \bibinfo {year} {1963})\ p.\ \bibinfo
  {pages} {100}\BibitemShut {NoStop}%
\bibitem [{\citenamefont {des Cloizeaux}(1965)}]{Cloizeaux}%
  \BibitemOpen
  \bibfield  {author} {\bibinfo {author} {\bibfnamefont {J.}~\bibnamefont {des
  Cloizeaux}},\ }\href@noop {} {\bibfield  {journal} {\bibinfo  {journal} {J.
  Phys. Chem. Solids}\ }\textbf {\bibinfo {volume} {26}},\ \bibinfo {pages}
  {259} (\bibinfo {year} {1965})}\BibitemShut {NoStop}%
\bibitem [{\citenamefont {Pomeranchuk}(1958)}]{pomeranchuk}%
  \BibitemOpen
  \bibfield  {author} {\bibinfo {author} {\bibfnamefont {I.}~\bibnamefont
  {Pomeranchuk}},\ }\href@noop {} {\bibfield  {journal} {\bibinfo  {journal}
  {Sov. Phys. JETP}\ }\textbf {\bibinfo {volume} {8}},\ \bibinfo {pages} {361}
  (\bibinfo {year} {1958})}\BibitemShut {NoStop}%
\bibitem [{\citenamefont {Keldysh}\ and\ \citenamefont
  {Kopaev}(1965)}]{Keldysh_Kopaev}%
  \BibitemOpen
  \bibfield  {author} {\bibinfo {author} {\bibfnamefont {L.~V.}\ \bibnamefont
  {Keldysh}}\ and\ \bibinfo {author} {\bibfnamefont {Y.~V.}\ \bibnamefont
  {Kopaev}},\ }\href@noop {} {\bibfield  {journal} {\bibinfo  {journal} {Sov.
  Phys. Solid State}\ }\textbf {\bibinfo {volume} {6}},\ \bibinfo {pages}
  {2219} (\bibinfo {year} {1965})}\BibitemShut {NoStop}%
\bibitem [{\citenamefont {Kozlov}\ and\ \citenamefont
  {Maksimov}(1965)}]{Kozlov_Maksimov}%
  \BibitemOpen
  \bibfield  {author} {\bibinfo {author} {\bibfnamefont {A.~N.}\ \bibnamefont
  {Kozlov}}\ and\ \bibinfo {author} {\bibfnamefont {L.~A.}\ \bibnamefont
  {Maksimov}},\ }\href@noop {} {\bibfield  {journal} {\bibinfo  {journal} {Sov.
  Phys. JETP}\ }\textbf {\bibinfo {volume} {21}},\ \bibinfo {pages} {790}
  (\bibinfo {year} {1965})}\BibitemShut {NoStop}%
\bibitem [{\citenamefont {Jerome}\ \emph {et~al.}(1967)\citenamefont {Jerome},
  \citenamefont {Rice},\ and\ \citenamefont {Kohn}}]{jerome_rice_kohn}%
  \BibitemOpen
  \bibfield  {author} {\bibinfo {author} {\bibfnamefont {D.}~\bibnamefont
  {Jerome}}, \bibinfo {author} {\bibfnamefont {T.~M.}\ \bibnamefont {Rice}}, \
  and\ \bibinfo {author} {\bibfnamefont {W.}~\bibnamefont {Kohn}},\ }\href@noop
  {} {\bibfield  {journal} {\bibinfo  {journal} {Phys. Rev.}\ }\textbf
  {\bibinfo {volume} {158}},\ \bibinfo {pages} {462} (\bibinfo {year}
  {1967})}\BibitemShut {NoStop}%
\bibitem [{\citenamefont {Halperin}\ and\ \citenamefont
  {Rice}(1968)}]{RMP_Halperin_rice}%
  \BibitemOpen
  \bibfield  {author} {\bibinfo {author} {\bibfnamefont {B.~I.}\ \bibnamefont
  {Halperin}}\ and\ \bibinfo {author} {\bibfnamefont {T.~M.}\ \bibnamefont
  {Rice}},\ }\href@noop {} {\bibfield  {journal} {\bibinfo  {journal} {Rev.
  Mod. Phys.}\ }\textbf {\bibinfo {volume} {40}},\ \bibinfo {pages} {755}
  (\bibinfo {year} {1968})}\BibitemShut {NoStop}%
\bibitem [{\citenamefont {Comte}\ and\ \citenamefont
  {Nozi\`{e}res}(1982)}]{Nozieres_Comte}%
  \BibitemOpen
  \bibfield  {author} {\bibinfo {author} {\bibfnamefont {C.}~\bibnamefont
  {Comte}}\ and\ \bibinfo {author} {\bibfnamefont {P.}~\bibnamefont
  {Nozi\`{e}res}},\ }\href@noop {} {\bibfield  {journal} {\bibinfo  {journal}
  {J. Phys. (Paris)}\ }\textbf {\bibinfo {volume} {43}},\ \bibinfo {pages}
  {1069} (\bibinfo {year} {1982})}\BibitemShut {NoStop}%
\bibitem [{\citenamefont {Hertz}(1976)}]{hertz}%
  \BibitemOpen
  \bibfield  {author} {\bibinfo {author} {\bibfnamefont {J.~A.}\ \bibnamefont
  {Hertz}},\ }\href@noop {} {\bibfield  {journal} {\bibinfo  {journal} {Phys.
  Rev. B}\ }\textbf {\bibinfo {volume} {14}},\ \bibinfo {pages} {1165}
  (\bibinfo {year} {1976})}\BibitemShut {NoStop}%
\bibitem [{\citenamefont {Abanov}\ and\ \citenamefont
  {Chubukov}(2004)}]{abanov_chubukov}%
  \BibitemOpen
  \bibfield  {author} {\bibinfo {author} {\bibfnamefont {A.}~\bibnamefont
  {Abanov}}\ and\ \bibinfo {author} {\bibfnamefont {A.}~\bibnamefont
  {Chubukov}},\ }\href@noop {} {\bibfield  {journal} {\bibinfo  {journal}
  {Phys. Rev. Lett.}\ }\textbf {\bibinfo {volume} {93}},\ \bibinfo {pages}
  {255702} (\bibinfo {year} {2004})}\BibitemShut {NoStop}%
\bibitem [{\citenamefont {v.~L\"{o}hneysen}\ \emph {et~al.}(2007)\citenamefont
  {v.~L\"{o}hneysen}, \citenamefont {Rosch}, \citenamefont {Vojta},\ and\
  \citenamefont {W\"{o}lfle}}]{RMP_hertz_breakdown}%
  \BibitemOpen
  \bibfield  {author} {\bibinfo {author} {\bibfnamefont {H.}~\bibnamefont
  {v.~L\"{o}hneysen}}, \bibinfo {author} {\bibfnamefont {A.}~\bibnamefont
  {Rosch}}, \bibinfo {author} {\bibfnamefont {M.}~\bibnamefont {Vojta}}, \ and\
  \bibinfo {author} {\bibfnamefont {P.}~\bibnamefont {W\"{o}lfle}},\
  }\href@noop {} {\bibfield  {journal} {\bibinfo  {journal} {Rev. Mod. Phys.}\
  }\textbf {\bibinfo {volume} {79}},\ \bibinfo {pages} {1015} (\bibinfo {year}
  {2007})}\BibitemShut {NoStop}%
\bibitem [{\citenamefont {Varma}\ \emph {et~al.}(2002)\citenamefont {Varma},
  \citenamefont {Nussinov},\ and\ \citenamefont {van
  Saarloos}}]{varma_SFL_review}%
  \BibitemOpen
  \bibfield  {author} {\bibinfo {author} {\bibfnamefont {C.~M.}\ \bibnamefont
  {Varma}}, \bibinfo {author} {\bibfnamefont {Z.}~\bibnamefont {Nussinov}}, \
  and\ \bibinfo {author} {\bibfnamefont {W.}~\bibnamefont {van Saarloos}},\
  }\href@noop {} {\bibfield  {journal} {\bibinfo  {journal} {Phys. Rep.}\
  }\textbf {\bibinfo {volume} {361}},\ \bibinfo {pages} {267} (\bibinfo {year}
  {2002})}\BibitemShut {NoStop}%
\bibitem [{\citenamefont {Varma}(1997)}]{Varma_1997}%
  \BibitemOpen
  \bibfield  {author} {\bibinfo {author} {\bibfnamefont {C.~M.}\ \bibnamefont
  {Varma}},\ }\href@noop {} {\bibfield  {journal} {\bibinfo  {journal} {Phys.
  Rev. B}\ }\textbf {\bibinfo {volume} {55}},\ \bibinfo {pages} {14554}
  (\bibinfo {year} {1997})}\BibitemShut {NoStop}%
\bibitem [{\citenamefont {Sun}\ and\ \citenamefont
  {Fradkin}(2008)}]{Fradkin_sun_TRV}%
  \BibitemOpen
  \bibfield  {author} {\bibinfo {author} {\bibfnamefont {K.}~\bibnamefont
  {Sun}}\ and\ \bibinfo {author} {\bibfnamefont {E.}~\bibnamefont {Fradkin}},\
  }\href@noop {} {\bibfield  {journal} {\bibinfo  {journal} {Phys. Rev. B}\
  }\textbf {\bibinfo {volume} {78}},\ \bibinfo {pages} {245122} (\bibinfo
  {year} {2008})}\BibitemShut {NoStop}%
\bibitem [{\citenamefont {Haldane}()}]{haldane_graphene_TRV}%
  \BibitemOpen
  \bibfield  {author} {\bibinfo {author} {\bibfnamefont {F.~D.~M.}\
  \bibnamefont {Haldane}},\ }\href@noop {} {\bibinfo  {journal} {Phys. Rev.
  Lett.}\ ,\ \bibinfo {pages} {2015}}\BibitemShut {NoStop}%
\bibitem [{\citenamefont {Raghu}\ \emph {et~al.}(2008)\citenamefont {Raghu},
  \citenamefont {Qi}, \citenamefont {Honerkamp},\ and\ \citenamefont
  {Zhang}}]{Zhang_TMI}%
  \BibitemOpen
\bibfield  {journal} {  }\bibfield  {author} {\bibinfo {author} {\bibfnamefont
  {S.}~\bibnamefont {Raghu}}, \bibinfo {author} {\bibfnamefont {X.-L.}\
  \bibnamefont {Qi}}, \bibinfo {author} {\bibfnamefont {C.}~\bibnamefont
  {Honerkamp}}, \ and\ \bibinfo {author} {\bibfnamefont {S.-C.}\ \bibnamefont
  {Zhang}},\ }\href@noop {} {\bibfield  {journal} {\bibinfo  {journal} {Phys.
  Rev. Lett.}\ }\textbf {\bibinfo {volume} {100}},\ \bibinfo {pages} {156401}
  (\bibinfo {year} {2008})}\BibitemShut {NoStop}%
\bibitem [{\citenamefont {Gavoret}\ \emph {et~al.}(1969)\citenamefont
  {Gavoret}, \citenamefont {Nozi\`{e}res}, \citenamefont {Roulet},\ and\
  \citenamefont {Combescot}}]{Nozieres_Gavoret}%
  \BibitemOpen
  \bibfield  {author} {\bibinfo {author} {\bibfnamefont {J.}~\bibnamefont
  {Gavoret}}, \bibinfo {author} {\bibfnamefont {P.}~\bibnamefont
  {Nozi\`{e}res}}, \bibinfo {author} {\bibfnamefont {B.}~\bibnamefont
  {Roulet}}, \ and\ \bibinfo {author} {\bibfnamefont {M.}~\bibnamefont
  {Combescot}},\ }\href@noop {} {\bibfield  {journal} {\bibinfo  {journal} {J.
  Phys. France}\ }\textbf {\bibinfo {volume} {30}},\ \bibinfo {pages} {987}
  (\bibinfo {year} {1969})}\BibitemShut {NoStop}%
\bibitem [{\citenamefont {Nozi\`{e}res}\ and\ \citenamefont
  {Comte}(1982)}]{Nozieres_Comte_2}%
  \BibitemOpen
  \bibfield  {author} {\bibinfo {author} {\bibfnamefont {P.}~\bibnamefont
  {Nozi\`{e}res}}\ and\ \bibinfo {author} {\bibfnamefont {C.}~\bibnamefont
  {Comte}},\ }\href@noop {} {\bibfield  {journal} {\bibinfo  {journal} {J.
  Phys. (Paris)}\ }\textbf {\bibinfo {volume} {43}},\ \bibinfo {pages} {1083}
  (\bibinfo {year} {1982})}\BibitemShut {NoStop}%
\bibitem [{\citenamefont {C\^{o}t\'{e}}\ and\ \citenamefont
  {Griffin}(1988)}]{cote_griffin}%
  \BibitemOpen
  \bibfield  {author} {\bibinfo {author} {\bibfnamefont {R.}~\bibnamefont
  {C\^{o}t\'{e}}}\ and\ \bibinfo {author} {\bibfnamefont {A.}~\bibnamefont
  {Griffin}},\ }\href@noop {} {\bibfield  {journal} {\bibinfo  {journal} {Phys.
  Rev. B}\ }\textbf {\bibinfo {volume} {37}},\ \bibinfo {pages} {4539}
  (\bibinfo {year} {1988})}\BibitemShut {NoStop}%
\bibitem [{\citenamefont {{S\'{a} de Melo}}\ \emph {et~al.}(1993)\citenamefont
  {{S\'{a} de Melo}}, \citenamefont {Randeria},\ and\ \citenamefont
  {Engelbrecht}}]{randeria_BCS_BEC}%
  \BibitemOpen
  \bibfield  {author} {\bibinfo {author} {\bibfnamefont {C.~A.~R.}\
  \bibnamefont {{S\'{a} de Melo}}}, \bibinfo {author} {\bibfnamefont
  {M.}~\bibnamefont {Randeria}}, \ and\ \bibinfo {author} {\bibfnamefont
  {J.~R.}\ \bibnamefont {Engelbrecht}},\ }\href@noop {} {\bibfield  {journal}
  {\bibinfo  {journal} {Phys. Rev. Lett.}\ }\textbf {\bibinfo {volume} {71}},\
  \bibinfo {pages} {3202} (\bibinfo {year} {1993})}\BibitemShut {NoStop}%
\bibitem [{\citenamefont {Loktev}\ \emph {et~al.}(2001)\citenamefont {Loktev},
  \citenamefont {Quick},\ and\ \citenamefont
  {Sharapov}}]{BCS_phase_Loktev_review}%
  \BibitemOpen
  \bibfield  {author} {\bibinfo {author} {\bibfnamefont {V.~M.}\ \bibnamefont
  {Loktev}}, \bibinfo {author} {\bibfnamefont {R.~M.}\ \bibnamefont {Quick}}, \
  and\ \bibinfo {author} {\bibfnamefont {S.~G.}\ \bibnamefont {Sharapov}},\
  }\href@noop {} {\bibfield  {journal} {\bibinfo  {journal} {Phys. Rep.}\
  }\textbf {\bibinfo {volume} {349}},\ \bibinfo {pages} {1} (\bibinfo {year}
  {2001})}\BibitemShut {NoStop}%
\bibitem [{\citenamefont {Balents}\ and\ \citenamefont
  {Varma}(2000)}]{varma_balents_X_PRL}%
  \BibitemOpen
  \bibfield  {author} {\bibinfo {author} {\bibfnamefont {L.}~\bibnamefont
  {Balents}}\ and\ \bibinfo {author} {\bibfnamefont {C.~M.}\ \bibnamefont
  {Varma}},\ }\href@noop {} {\bibfield  {journal} {\bibinfo  {journal} {Phys.
  Rev. Lett.}\ }\textbf {\bibinfo {volume} {84}},\ \bibinfo {pages} {1264}
  (\bibinfo {year} {2000})}\BibitemShut {NoStop}%
\bibitem [{\citenamefont {Barzykin}\ and\ \citenamefont
  {Gorkov}(2000)}]{X_Gorkov}%
  \BibitemOpen
  \bibfield  {author} {\bibinfo {author} {\bibfnamefont {V.}~\bibnamefont
  {Barzykin}}\ and\ \bibinfo {author} {\bibfnamefont {L.~P.}\ \bibnamefont
  {Gorkov}},\ }\href@noop {} {\bibfield  {journal} {\bibinfo  {journal} {Phys.
  Rev. Lett.}\ }\textbf {\bibinfo {volume} {84}},\ \bibinfo {pages} {2207}
  (\bibinfo {year} {2000})}\BibitemShut {NoStop}%
\bibitem [{\citenamefont {Nambu}\ and\ \citenamefont
  {Jona-Lasinio}(1961{\natexlab{a}})}]{NJL_1}%
  \BibitemOpen
  \bibfield  {author} {\bibinfo {author} {\bibfnamefont {Y.}~\bibnamefont
  {Nambu}}\ and\ \bibinfo {author} {\bibfnamefont {G.}~\bibnamefont
  {Jona-Lasinio}},\ }\href@noop {} {\bibfield  {journal} {\bibinfo  {journal}
  {Phys. Rev.}\ }\textbf {\bibinfo {volume} {122}},\ \bibinfo {pages} {345}
  (\bibinfo {year} {1961}{\natexlab{a}})}\BibitemShut {NoStop}%
\bibitem [{\citenamefont {Nambu}\ and\ \citenamefont
  {Jona-Lasinio}(1961{\natexlab{b}})}]{NJL_2}%
  \BibitemOpen
  \bibfield  {author} {\bibinfo {author} {\bibfnamefont {Y.}~\bibnamefont
  {Nambu}}\ and\ \bibinfo {author} {\bibfnamefont {G.}~\bibnamefont
  {Jona-Lasinio}},\ }\href@noop {} {\bibfield  {journal} {\bibinfo  {journal}
  {Phys. Rev.}\ }\textbf {\bibinfo {volume} {124}},\ \bibinfo {pages} {246}
  (\bibinfo {year} {1961}{\natexlab{b}})}\BibitemShut {NoStop}%
\bibitem [{\citenamefont {Popov}(1987)}]{book_popov}%
  \BibitemOpen
  \bibfield  {author} {\bibinfo {author} {\bibfnamefont {V.~N.}\ \bibnamefont
  {Popov}},\ }\href@noop {} {\emph {\bibinfo {title} {Functional Integrals and
  Collective Excitations}}}\ (\bibinfo  {publisher} {Cambridge Univ. Press,
  Cambridge},\ \bibinfo {year} {1987})\BibitemShut {NoStop}%
\bibitem [{\citenamefont {Kleinert}(1978)}]{book_kleinart}%
  \BibitemOpen
  \bibfield  {author} {\bibinfo {author} {\bibfnamefont {H.}~\bibnamefont
  {Kleinert}},\ }\href@noop {} {\bibfield  {journal} {\bibinfo  {journal}
  {Fortsch. Phys.}\ }\textbf {\bibinfo {volume} {26}},\ \bibinfo {pages} {565}
  (\bibinfo {year} {1978})}\BibitemShut {NoStop}%
\bibitem [{\citenamefont {de~Gennes}(1963)}]{deGennes_ising}%
  \BibitemOpen
  \bibfield  {author} {\bibinfo {author} {\bibfnamefont {P.~G.}\ \bibnamefont
  {de~Gennes}},\ }\href@noop {} {\bibfield  {journal} {\bibinfo  {journal}
  {Solid State Commun.}\ }\textbf {\bibinfo {volume} {1}},\ \bibinfo {pages}
  {132} (\bibinfo {year} {1963})}\BibitemShut {NoStop}%
\bibitem [{\citenamefont {Brout}\ \emph {et~al.}(1966)\citenamefont {Brout},
  \citenamefont {M\"{u}ller},\ and\ \citenamefont {Thomas}}]{brout_ising}%
  \BibitemOpen
  \bibfield  {author} {\bibinfo {author} {\bibfnamefont {R.}~\bibnamefont
  {Brout}}, \bibinfo {author} {\bibfnamefont {K.~A.}\ \bibnamefont
  {M\"{u}ller}}, \ and\ \bibinfo {author} {\bibfnamefont {H.}~\bibnamefont
  {Thomas}},\ }\href@noop {} {\bibfield  {journal} {\bibinfo  {journal} {Solid
  State Commun.}\ }\textbf {\bibinfo {volume} {4}},\ \bibinfo {pages} {507}
  (\bibinfo {year} {1966})}\BibitemShut {NoStop}%
\bibitem [{\citenamefont {Lipkin}(2002)}]{book_lipkin}%
  \BibitemOpen
  \bibfield  {author} {\bibinfo {author} {\bibfnamefont {H.~J.}\ \bibnamefont
  {Lipkin}},\ }\href@noop {} {\emph {\bibinfo {title} {Lie Groups for
  Pedestrians}}}\ (\bibinfo  {publisher} {Dover Publications, New York},\
  \bibinfo {year} {2002})\BibitemShut {NoStop}%
\bibitem [{\citenamefont {Abrikosov}\ \emph {et~al.}(1975)\citenamefont
  {Abrikosov}, \citenamefont {Gorkov},\ and\ \citenamefont
  {Dzyaloshinski}}]{bluebook}%
  \BibitemOpen
  \bibfield  {author} {\bibinfo {author} {\bibfnamefont {A.~A.}\ \bibnamefont
  {Abrikosov}}, \bibinfo {author} {\bibfnamefont {L.~P.}\ \bibnamefont
  {Gorkov}}, \ and\ \bibinfo {author} {\bibfnamefont {I.~E.}\ \bibnamefont
  {Dzyaloshinski}},\ }\href@noop {} {\emph {\bibinfo {title} {Methods of
  Quantum Field Theory in Statistical Physics}}}\ (\bibinfo  {publisher} {Dover
  Publications, New York},\ \bibinfo {year} {1975})\BibitemShut {NoStop}%
\bibitem [{\citenamefont {Volkov}\ \emph {et~al.}(1976)\citenamefont {Volkov},
  \citenamefont {Kopaev},\ and\ \citenamefont {Rusinov}}]{FM_volkov}%
  \BibitemOpen
  \bibfield  {author} {\bibinfo {author} {\bibfnamefont {B.~A.}\ \bibnamefont
  {Volkov}}, \bibinfo {author} {\bibfnamefont {Y.~V.}\ \bibnamefont {Kopaev}},
  \ and\ \bibinfo {author} {\bibfnamefont {A.~I.}\ \bibnamefont {Rusinov}},\
  }\href@noop {} {\bibfield  {journal} {\bibinfo  {journal} {Sov. Phys. JETP}\
  }\textbf {\bibinfo {volume} {41}},\ \bibinfo {pages} {952} (\bibinfo {year}
  {1976})}\BibitemShut {NoStop}%
\bibitem [{\citenamefont {Tinkham}(1992)}]{tinkham_book}%
  \BibitemOpen
  \bibfield  {author} {\bibinfo {author} {\bibfnamefont {M.}~\bibnamefont
  {Tinkham}},\ }\href@noop {} {\emph {\bibinfo {title} {Group Theory and
  Quantum Mechanics}}}\ (\bibinfo  {publisher} {Dover Publications, New York},\
  \bibinfo {year} {1992})\BibitemShut {NoStop}%
\bibitem [{\citenamefont {Ma{\~{n}}es}\ \emph {et~al.}(2007)\citenamefont
  {Ma{\~{n}}es}, \citenamefont {Guinea},\ and\ \citenamefont
  {Vozmediano}}]{IT_graphene}%
  \BibitemOpen
  \bibfield  {author} {\bibinfo {author} {\bibfnamefont {J.~L.}\ \bibnamefont
  {Ma{\~{n}}es}}, \bibinfo {author} {\bibfnamefont {F.}~\bibnamefont {Guinea}},
  \ and\ \bibinfo {author} {\bibfnamefont {M.~A.~H.}\ \bibnamefont
  {Vozmediano}},\ }\href@noop {} {\bibfield  {journal} {\bibinfo  {journal}
  {Phys. Rev. B}\ }\textbf {\bibinfo {volume} {75}},\ \bibinfo {pages} {155424}
  (\bibinfo {year} {2007})}\BibitemShut {NoStop}%
\bibitem [{\citenamefont {Varma}(1999)}]{varma_1999}%
  \BibitemOpen
  \bibfield  {author} {\bibinfo {author} {\bibfnamefont {C.~M.}\ \bibnamefont
  {Varma}},\ }\href@noop {} {\bibfield  {journal} {\bibinfo  {journal} {Phys.
  Rev. Lett.}\ }\textbf {\bibinfo {volume} {83}},\ \bibinfo {pages} {3538}
  (\bibinfo {year} {1999})}\BibitemShut {NoStop}%
\bibitem [{\citenamefont {Singh}\ and\ \citenamefont
  {Te\v{s}anovi\'{c}}(1990)}]{AFM_Singh_1}%
  \BibitemOpen
  \bibfield  {author} {\bibinfo {author} {\bibfnamefont {A.}~\bibnamefont
  {Singh}}\ and\ \bibinfo {author} {\bibfnamefont {Z.}~\bibnamefont
  {Te\v{s}anovi\'{c}}},\ }\href@noop {} {\bibfield  {journal} {\bibinfo
  {journal} {Phys. Rev. B}\ }\textbf {\bibinfo {volume} {41}},\ \bibinfo
  {pages} {614} (\bibinfo {year} {1990})}\BibitemShut {NoStop}%
\bibitem [{\citenamefont {Singh}\ \emph {et~al.}(1991)\citenamefont {Singh},
  \citenamefont {Te\v{s}anovi\'{c}},\ and\ \citenamefont {Kim}}]{AFM_Singh_2}%
  \BibitemOpen
  \bibfield  {author} {\bibinfo {author} {\bibfnamefont {A.}~\bibnamefont
  {Singh}}, \bibinfo {author} {\bibfnamefont {Z.}~\bibnamefont
  {Te\v{s}anovi\'{c}}}, \ and\ \bibinfo {author} {\bibfnamefont {J.~H.}\
  \bibnamefont {Kim}},\ }\href@noop {} {\bibfield  {journal} {\bibinfo
  {journal} {Phys. Rev. B (R)}\ }\textbf {\bibinfo {volume} {44}},\ \bibinfo
  {pages} {7757} (\bibinfo {year} {1991})}\BibitemShut {NoStop}%
\end{thebibliography}

%

\end{document}